\documentclass[12pt,preprint]{aastex}
\begin{document}

\title{The Young Cluster in IC\,1274}

\author{S. E. Dahm\altaffilmark{1}, G. H. Herbig\altaffilmark{2}, \& Brendan P. Bowler\altaffilmark{2}}

\altaffiltext{1}{W. M. Keck Observatory, 65-1120 Mamalahoa Hwy, Kamuela, HI 96743, USA}
\altaffiltext{2}{Institute for Astronomy, University of Hawaii, 2680 Woodlawn Drive, Honolulu, HI 96822, USA}

\begin{abstract}
IC\,1274 is a faintly luminous nebula lying on the near surface of the Lynds 227 
(L227) molecular cloud. Its cavity-like morphology is reminiscent of a blistered star-forming 
region. Four luminous, early-type (B0--B5) stars are located within a spherical volume 
$\sim$5\arcmin\ in diameter that appears to be clear of heavy obscuration. Approximately
centered in the cleared region is the B0 V star HD 166033, which is thought to be largely
responsible for the cavity's excavation. Over 80 H$\alpha$ emission sources brighter 
than $V\sim21$ have been identified in the region. More than half of these are concentrated
in IC\,1274 and are presumably members of a faint T Tauri star population. {\it Chandra} 
Advanced CCD Imaging Spectrometer 
(ACIS) imaging of a nearby suspected pulsar and time-variable $\gamma$-ray source (GeV J1809-2327)
detected 21 X-ray sources in the cluster vicinity, some of which are coincident with
the early-type stars and H$\alpha$ emitters in IC\,1274. Deep ($V\sim22$) optical $BVRI$ 
photometry has been obtained for the cluster region. A distance of $1.82 \pm 0.3$ kpc and a mean 
extinction of $A_{V}\sim1.21\pm0.2$ mag follow from photometry of the early-type stars.
Using pre-main-sequence evolutionary models, we derive a median age for the H$\alpha$ emitters
and X-ray sources of $\sim$1 Myr; however, a significant dispersion is present. Our interpretation 
of the structure of IC\,1274 and the spatial distribution of H$\alpha$ emitters is that the 
early-type stars formed recently and are in the process
of ionizing and dispersing the molecular gas on the near 
surface of L227. The displaced material was driven against what remains of the molecular
cloud to the east, enabling the formation of the substantial number of T Tauri stars found 
there. A dispersed population of H$\alpha$ emitters is also found along the periphery 
of L227, IC\,1275, and IC\,4684. These sources, if pre-main-sequence stars, appear to have
formed in relative isolation compared to the dense cluster environment of IC\,1274 or,
alternatively, may be older and have drifted further from their formation sites. We
identify a $V\sim21.5$ star very near the position of X-ray source 5, the assumed $\gamma$-ray source
and young pulsar candidate. The lack of distinctive characteristics for this source,
however, coupled with the density of faint stars in this region suggest that this may
be a random superposition.
\end{abstract}


\section{Introduction}

IC\,1274 ($\alpha_{J2000}=18^{\rm h}09^{\rm m}47^{\rm s}$, $\delta=-23{\arcdeg}38\arcmin45\arcsec$; $l,b=7\fdg3$, ${-2}\fdg0$),
shown in Figure 1, lies within an aggregate of bright nebulae located ${\sim}1\fdg5$ northeast of the \ion{H}{2}
region M8 and nearly 2$\arcdeg$ southeast of M20. Optical images show that M8 is located at the 
base of a bulbous mass of dark nebulosity, $\sim$0\fdg8 wide, that appears to project from 
the Galactic plane to a distance of $\sim$1\fdg7 before terminating in a complex region 
referred to as Simeis 188. Included within Simeis 188 are the bright nebulae IC\,4684, IC\,1274, 
IC\,1275, IC\,4685, and NGC\,6559, as well as the Lynds 227 (L227) and Lynds 210 (L210) molecular clouds. 
Low-level \ion{H}{2} emission present throughout the region is referred to as Sharpless 2-32 
(Sh 2-32). Shown in Figure 2 is a 45\arcmin$\times$56\arcmin\ blue image of Simeis 188 obtained 
from the Digitized Sky Survey in which major components are identified. IC\,1274 is shown in more
detail in Figure 3, a deep $R$-band image obtained with the University of Hawaii (UH) 2.2 m telescope.
Four bright B-type stars, which are labeled, appear to illuminate the nebula. A spherical cavity
$\sim$5\arcmin\ in diameter appears to have been carved out of L227, presumably by the 
early-type stars.

Approximately centered in the evacuated region is HD 166033 (B0 V). This star likely furnishes the 
illumination of the bright nebular rim, although the confused structure of the rim near 
CoD$-$23$^{\circ}$13997 (B1 V) suggests that this star is also involved. Together with HD 314032 (B5 V) and 
HD 314031 (B0.5 V), these stars could represent the most massive members of a young stellar cluster that 
is emerging from the molecular cloud. Herbig (1957) identified six faint H$\alpha$ emission stars in 
and around IC\,1274 using the slitless grating spectrograph of the Crossley reflector at Lick Observatory. 
To date, no thorough investigation of the young stellar population of IC\,1274 is known to the authors.

The region received considerable attention when the Energetic Gamma Ray Experiment Telescope (EGRET)
onboard the {\it Compton Gamma Ray Observatory} discovered an apparently time-variable $\gamma$-ray 
source, GeV J1809-2327 (Oka et al.\ 1999; Hartman et al.\ 1999; Braje et al.\ 2002), several arcminutes
north of IC\,1274. Subsequent X-ray observations of the region with {\it Advanced Satellite for Cosmology and
Astrophysics} {\it (ASCA)} and {\it Chandra} 
identified 21 X-ray sources, many of which correspond to stars in the IC\,1274 region, including 
the four early-type stars. Braje et al.\ (2002) give particular attention to one source that is attached 
to a 30\arcsec\ tail of diffuse X-ray emission. They speculate that this source is a young pulsar with 
no optical counterpart and the likely origin of the $\gamma$-ray emission. 

The present investigation is intended to identify the low-mass population of the star-forming region. 
Toward this end, deep ($V\sim22$) optical ($BVRI$) imaging of IC\,1274 has been obtained using the 
Large Format Camera (LFC) on the Hale 200 inch (5 m) telescope at Palomar Observatory. Complementing these observations are three epochs 
of slitless grism H$\alpha$ spectroscopy spanning some $\sim$19 yr that reveal a faint T Tauri 
star (TTS) population concentrated within the cluster and sparsely distributed along the
periphery of the L227 molecular cloud. High-dispersion optical spectra of the four early-type stars
in IC\,1274 and of three regions within the weakly ionized emission nebula have also been obtained.

This paper is organized as follows: In \S\,2 details are provided of all observations presented.
In \S\,3 distance and extinction estimates are determined for the region and compared with results
from the literature for Sh 2-32. In \S\,4 
the H$\alpha$ slitless grism survey of IC\,1274 and the surrounding area is discussed, potential 
sources of contamination are reviewed, and the spatial distributions of detected H$\alpha$ emitters 
and X-ray sources are considered. In 
 \S\,5, we examine the cluster, including the color-magnitude diagram, the ages and masses of the
identified TTSs and X-ray sources, and its near-infrared color--color diagram. In \S\S\,6 and 7, 
the high-dispersion spectra of the early-type stars and nebulosity are discussed. The electron density 
is also derived for the luminous shell. Finally, we discuss star formation in IC\,1274, the young 
pulsar candidate identified by Braje et al.\ (2002) and the star-forming history of the region.

\section{Observations}

\subsection{Optical Photometry}
Deep $BVRI$ (Bessel) images of IC\,1274 were obtained on the night of 2006 August 31 using the LFC 
mounted at the prime focus of the Hale telescope. The night was photometric; however, 
seeing conditions varied between ${\sim}1\farcs0$ and 1\farcs5. The LFC is a mosaic of six SITe 2048$\times$4096
pixel thinned, backside-illuminated, anti-reflection coated CCDs with $\sim$15\arcsec\ gaps between 
detectors. The mosaic yields a $\sim$24\arcmin\ diameter field of view with a pixel scale of 0.36\arcsec\ 
pixel$^{-1}$ (binned 2$\times$2). A sequence of five dithered images was obtained with exposure times of 
120 s in the $B$- and $V$-band filters and 60 s in the $R$- and $I$-band filters. A second sequence 
with 10 s and 5 s integration times was also made with the $B$- and $V$-band filters and the $R$- and 
$I$-band filters, respectively, to enable photometry on stars saturated in the longer integrations. The 
five dithers allowed regions lying in the detector gaps to be adequately covered. 

The reduction of the LFC mosaic imaging followed a description by Roy Gal that makes use of the external 
packages $mscred$ and $mscdb$ in the Image Reduction and Analysis Facility (IRAF). The $ccdproc$ task 
was used to apply crosstalk and overscan corrections, and trim the overscan regions of the individual 
CCDs. The combined bias frame and appropriate dome flats were then applied to the standard star and 
target images yielding processed data frames. Remnant large-scale background variations caused by 
uneven dome illumination were removed by creating and applying sky flats generated from twilight images. 
Due to significant distortion present in the LFC, corrections to the astrometric solutions must be 
derived and applied using the $mscsetwcs$ task, which matches catalogs of USNO-A2 stars to each of the 
observed fields. Stars were rejected on the individual frames until residuals for a 5$^{th}$ order 
polynomial fit were $<$0\farcs4. Cosmic rays were removed using the $craverage$ task in the $crutil$ 
package of IRAF. Given that $craverage$ does not function on multiextension format (MEF) images, each 
CCD in the mosaic was treated individually. 

Single images were created from the MEF files using 
the $mscimage$ task. Images of a given field were then tangent plane projected with the same pixel scale 
and center position and stacked into a single image using the $mscstack$ task in IRAF. Aperture photometry 
was then performed on the final stacked images using the $phot$ task in the $daophot$ package using 
apertures of 1\farcs8 for both standard stars and program stars. Extinction corrections and transformation 
to the Landolt (1992) system were achieved by observing several standard star fields throughout the 
night at various air masses. Density plots of photometric errors as they depend upon $V$-band mag are 
shown for $V$, $V-R$, and $V-I$ in Figure 4.

\subsection{H$\alpha$ Slitless Grism Spectroscopy}
The H$\alpha$ emission surveys of the IC\,1274 region were carried out on 1990 September 14, 2003 October 02--03, 
and 2009 July 17 UT. The 1990 and 2003 surveys were made with the Wide-Field Grism Spectrograph (WFGS)
installed at the f/10 Cassegrain focus of the UH 2.2 m telescope.
A 420 line mm$^{-1}$ grism blazed at 6400 \AA\ provided a dispersion of 3.85 \AA\ pixel$^{-1}$. 
The narrowband H$\alpha$ filter isolated a region of the first-order spectra between $\sim$6300 and 6750 \AA.
WFGS imaged upon the central 1024$\times$1024 pixels of a Tektronix 2048$\times$2048 pixel CCD, yielding
a nominal field of view $\sim$5\farcm5$\times$5\farcm5. The 1990 observations covered a region 
$\sim$15\arcmin$\times$5\farcm5 in area, roughly centered upon HD 166033. The more extensive 2003 
survey consisted of nine overlapping fields arranged in a 3$\times$3 grid pattern yielding a final mapped
region $\sim$15\arcmin$\times$15\arcmin\ in area. Integration times for the slitless grism spectra were 30, 120, 
and 1200 s for each field to ensure complete, unsaturated coverage for nearly all sources. Seeing conditions 
during the 2003 October  survey were excellent, $\le$0\farcs6. Consequently, several faint H$\alpha$ emitters 
were only detected by those observations.

The 2009 survey used the Wide-Field Grism Spectrograph 2 (WFGS2: Uehara et al.\ 2004) also installed at the Cassegrain focus
of the UH 2.2 m telescope. The WFGS2 observations used a 300 line mm$^{-1}$ grism blazed at 6500 \AA\
providing a dispersion of 3.8 \AA\ pixel$^{-1}$. The narrowband H$\alpha$ filter has a 500 \AA\
passband centered near 6515 \AA. The detector for WFGS2 is a Tektronix 2048$\times$2048 CCD with 24 $\mu$m 
pixels. The field of view is $\sim$11\farcm5$\times$11\farcm5. The 2009 survey was the most extensive in area,
covering a region $\sim$30\arcmin$\times$30\arcmin\ on the sky, but the deepest integrations were limited 
to 240 s due to telescope guiding problems. Depending upon seeing conditions, the limiting
measurable equivalent width, W(H$\alpha$), is approximately 2 \AA\ for both WFGS and WFGS2. The continua 
of stars brighter than $V\sim21.0$ were sufficiently well defined on deep exposures such that 
W(H$\alpha$) could be determined.

\subsection{HIRES Spectroscopy}
High-dispersion optical spectra were obtained for the four early-type stars in IC\,1274 and across three regions
of the nebulosity using the High Resolution Echelle Spectrometer (HIRES; Vogt et al.\ 1994) on Keck I on
the night of 2010 June 16. Conditions were photometric with average seeing of $\sim$1\farcs0. 
High-resolution spectra were also obtained for HD 166033 on 1997 August 12 and 1998 October 31 using HIRES.
The instrument was configured with the red cross-disperser and collimator in beam. The C1 decker 
(0\farcs87$\times$7\farcs0), which has a projected slit width of 3 pixels, was used, providing
a spectral resolution of $\sim$45,000 ($\sim$6.7 km s$^{-1}$). Near complete spectral coverage from
$\sim$3600--8000 \AA\ was achieved. The 1997 and 1998 observations were made with the single Tektronix
2048$\times$2048 CCD that has 24 $\mu$m pixels. The 2010 observations were made with the upgraded
3-chip mosaic of MIT-LL CCDs having 15 $\mu$m pixels. The CCDs were used in low-gain mode, resulting
in readout noise levels of $\sim$2.8, 3.1, and 3.1 e$^{-1}$ for the red, green, and blue detectors,
respectively. Internal quartz lamps were used for flat fielding, and ThAr lamp spectra were obtained
for wavelength calibration. Integration times were 900 s, yielding signal-to-noise levels of $\sim$100
on the green detector for the early-type stars. The cross-dispersed spectra were reduced and extracted
using standard tools available in IRAF. 

\section{Distance and Extinction}
Direct imaging of the Sagittarius star-forming region suggests that Simeis 188, M8, and M20 are associated
with the same complex of molecular gas and dust (Herbig 1957). Modern distance estimates for these 
star-forming regions and their associated clusters include 1247 pc for NGC\,6530 (Mayne \& Naylor 2008), 1250 pc
for M8 (Arias et al.\ 2007), 1260 pc for NGC\,6531 (Park et al.\ 2001), and 1670 pc for M20 (Rho et al.\ 2008).
A recent study of variations in the extinction law within the Trifid nebula (M20) by Cambr{\'e}sy et al.\ (2011),
however, suggests an even greater distance of 2.7$\pm$0.5 kpc. The extinction values were obtained using
Two Micron All Sky Survey (2MASS), United Kingdom Infrared Deep Sky Survey (UKIDSS) and {\it Spitzer Space
Telescope} data, but are insensitive to the diffuse interstellar medium, which is characterized by a ratio
of total-to-selective absorption of $R_{V}=3.1$.

To estimate the distance of IC\,1274, we assume that the four early-type stars define the zero-age
main sequence. These stars, however, are saturated even on the shortest LFC integrations. We therefore use
the optical photometry for these stars from Herbst et al.\ (1982). We also adopt the $M_{V}$ magnitudes and
intrinsic $B-V$ colors for main-sequence stars of Schmidt-Kaler (1982) to determine individual $A_{V}$ values
and distances, assuming $R_{V}=3.1$, which are presented in Table 1. These $M_{V}$ values rely upon spectral 
types taken from Herbst et al.\ (1982) and Crampton \& Fisher (1974). No modern spectral types for these stars
are available. The average distance is $1.82\pm0.3$ kpc 
and the mean $A_{V}$ is ${\sim}1.21\pm0.2$ mag. One caveat regarding this distance estimate is that 
Jaschek \& Gomez (1998) find that for a given B-type dwarf, $M_{V}$ is distributed around its mean with a 
standard deviation of 0.55 mag. For the sample of four stars in IC\,1274, we expect an inherent error of
0.55/4$^{1/2}$ mag, which equates to distance uncertainty of $\sim$245 pc.

This distance estimate is consistent with those derived for the surrounding \ion{H}{2} region Sh 2-32 by other
investigations using a range of techniques. Georgelin et al.\ (1973) derive $UBV$ photometric and H$\alpha$ 
interferometric kinematic distances of 2.4 and 1.88$\pm$1.03 kpc, respectively. From $UBV$ and H$\beta$ 
photoelectric photometry of 5 early-type stars within Sh 2-32, Vogt \& Moffat (1975) determine a distance of 
2.2 kpc and a mean extinction of $A_{V}\sim$1.17 mag. Using CO velocities for molecular gas within Sh 2-32,
Fich \& Blitz (1984) find a kinematic distance of 1.80$\pm$0.6 kpc. Photometric distance estimates are sensitive to
errors in spectral classification and luminosity type, absolute magnitude, binarity, interstellar absorption, and
measurement errors. Kinematic distances are limited by the accuracy of the Galactic rotation curve and the assumption 
of circular motion about the Galactic center. Random motions, expansion of the \ion{H}{2} region, and 
measurement errors contribute to the total uncertainty as well. Table 2 summarizes these results from the literature, as well as
the methods used in their determination.

\section{The H$\alpha$ Emission Survey of IC\,1274 and L227}

\subsection{Detected H$\alpha$ Emission Sources and Field Star Contamination}
Herbig (1957) discovered six H$\alpha$ emission stars in and around IC\,1274 (LkH$\alpha$125--130) in a
photographic survey carried out at Lick Observatory between 1954 and 1956. The limiting magnitude of
the Lick survey was $R\sim17$. The H$\alpha$ slitless grism observations presented here resulted in
the detection of 82 H$\alpha$ emitters, including 5 of those identified by Herbig (1957), in the 
30\arcmin$\times$30\arcmin\ survey area. Over half of these emitters are concentrated within IC\,1274.
The remaining sources are distributed across the region, but are preferentially found along the
periphery of L227.

Provided in Table 3 are the J2000 coordinates, $V$-band magnitudes, $V-R$, $V-I$, $J-H$, $H-K_{S}$ 
colors, and $K_{S}$-band magnitudes for these emission-line sources, if available. The imaging region 
with the LFC was less extensive than the H$\alpha$ emission survey; consequently some of the 
detected H$\alpha$ emitters lack optical photometry. Those sources lacking near-infrared photometry 
are either fainter than the detection limit of 2MASS, or their point-spread functions are confused 
with other nearby sources. Also given in Table 3 are the measured W(H$\alpha$) values from each of the three 
H$\alpha$ surveys (1990, 2003, 2009), if available. For the previously unknown emission-line sources, 
an IH$\alpha$ number is assigned, continuing the numbering convention of Herbig (1998), Herbig \&
Dahm (2002), Herbig et al.\ (2004), Dahm \& Simon (2005), Dahm (2005), and Herbig \& Dahm (2006). 
No corrections have been applied to the W(H$\alpha$) values presented in Table 3 for overlying 
absorption structure.

The fundamental advantage of spectroscopic H$\alpha$ surveys is the near unambiguous detection
of pre-main-sequence stars in young clusters and star-forming regions. The primary sources of
contamination are active late-K and M-type field stars (dMe) that exhibit enhanced chromospheric
and coronal emission. H$\alpha$ emission strengths among dMe stars, however, are typically
weak, W(H$\alpha$) $<$ 10 \AA\ (Hodgkin et al.\ 1995; Reid et al.\ 1995; Hawley et al.\ 1996), and
would predominantly affect the statistics of weak-line T Tauri star (WTTS) populations.
Few (19/82) weak-line emitters are detected among the H$\alpha$ emission stars in this survey.
Other potential sources of contamination include chromospherically active giants, RS CVn binaries,
cataclysmic variables, and active galaxies. The field density of such objects is expected to be
low, and the L227 molecular cloud would likely obscure background sources over a substantial fraction
of the survey area. With these caveats, we assume that field contamination among the H$\alpha$ 
emission stars is low, but certainly nonzero. Most of the detected emitters are believed to be
members of a faint TTS population.

The equivalent width of H$\alpha$ is a well-established indicator of accretion processes and 
chromospheric activity among pre-main-sequence stars (for a review, see Bertout 1989). 
Traditionally, the boundary between classical T Tauri stars (CTTS) and WTTS was placed at
W(H$\alpha$) = 10 \AA\ (e.g. Herbig 1998). No physical interpretation was intended for this value,
but clear differences in the processes responsible for emission have since been recognized
for CTTS (accretion) and WTTS (enhanced chromospheric activity). 
Because of the contrast effect (Basri \& Marcy 1995; White \& Basri 2003), W(H$\alpha$)
is spectral-type dependent. Consequently, no unique W(H$\alpha$) value is capable of distinguishing
all accretors from nonaccretors. Various spectral type dependent criteria have been
proposed, (e.g., Mart\'in 1998; White \& Basri 2003), but without spectral type information for
the low-mass population of IC\,1274, we assume the classical boundary of 10 \AA\ to distinguish
CTTS from WTTS.

Using this criterion, we find the fraction of H$\alpha$ emitters that are WTTS, 
$f$(WTTS) = $N$(WTTS)/$N$(TTS), to be $0.23\pm0.07$. We compare this fraction with
those found in other young clusters and star-forming regions derived from identical
techniques. In IC\,5146 this ratio is $0.23 \pm 0.06$ (Herbig \& Dahm 2002); in
NGC\,2264, 0.43$\pm$0.03 (Dahm \& Simon 2005); in IC\,348, $0.52\pm0.12$ Herbig (1998);
and in NGC\,2362, $0.91\pm0.08$ (Dahm 2005). The published ages for these young clusters are
$\sim$1.0 Myr for IC\,5146 (Herbig \& Dahm 2002), $\sim$1--2.5 Myr for NGC\,2264 (Dahm \& Simon 2005; 
Dahm et al.\ 2007), $\sim$1--3 Myr for IC\,348 (Lada et al.\ 2006), and $\sim$5 Myr for NGC\,2362
(Dahm 2005; Dahm \& Hillenbrand 2007). The decay of strong H$\alpha$ emission with isochronal
age parallels the timescale of inner disk dissipation determined from near- and mid-infrared 
excess observations (e.g., Haisch et al.\ 2001; Hern{\'a}ndez et al.\ 2007).

Notably absent from IC\,1274 are bright H$\alpha$ emitters that could be intermediate-mass Herbig
AeBe stars or classical Be stars whose circumstellar disks have become optically thin (Hillenbrand 
et al.\ 1993). Given the relative youth of the cluster, the presence of such stars would be 
expected. Further north near IC\,4684 is one potential intermediate-mass emission-line star: 
LkH$\alpha$127. Optical photometry is available from the literature, $V=11.81$ mag. Assuming the
distance and mean extinction of IC\,1274, the $M_{V}$ for this source would be $\sim-0.7$ mag, 
consistent with a mid-B spectral type at $\sim$1 Myr using the isochrones of Siess et al.\ (2000).
A spectrum, however, is needed for confirmation.

\subsection{Spatial Distribution of H$\alpha$ Emitters and X-ray Sources}
Shown in Figure 5 is the spatial distribution of H$\alpha$ emission sources superimposed upon a
red Digitized Sky survey image of the IC\,1274 region. Also plotted are the X-ray sources from 
Braje et al.\ (2002) that were detected in a shallow (9.7 ks) {\it Chandra} ACIS exposure (ObsID 739). The four early-type
stars, as well as four H$\alpha$ emitters, were detected in X-rays. The other X-ray sources 
lacking H$\alpha$ emission are presumably cluster members. Two are not detected optically and
two others lie far outside of the LFC field of view (annotated in Fig.\ 5). In Table 4 we summarize the properties of 
the 21 X-ray sources identified by Braje et al.\ (2002), including their J2000 coordinates, 
$V$-band magnitudes, $V-R$, $V-I$, $J-H$, $H-K_{S}$ colors, and $K_{S}$-band magnitudes.

Across the entire region surveyed, roughly half of the detected H$\alpha$ emitters and X-ray 
sources are located in IC\,1274. Most of these are preferentially found in the southeastern 
hemisphere of the evacuated cavity. Faint sources, some of which are H$\alpha$ emitters, also
appear to be strongly concentrated around HD 166033, the assumed center of IC\,1274.
This is not observed for the other B-type
stars. Shown in Figures 6a and 6b are $I$-band images of HD 166033 obtained with
the UH 2.2 m telescope. They reveal at least a dozen sources within $\sim$20\arcsec\ of HD 166033,
or projected separations of $\le$0.18 pc. Considering only H$\alpha$ emitters and X-ray sources,
we estimate the surface density of presumed cluster members to peak $\sim$8 pc$^{-2}$ within the
central $\sim$1 pc, comparable to other densely populated star-forming regions.

It is perhaps no coincidence that IC\,1274 hosts the most massive star that has yet formed
in the L227 molecular cloud. The dense population of H$\alpha$ emitters detected in the
central cluster region suggests that star formation in IC\,1274 was highly localized and
rapid as random velocity dispersions on the order of 1 km s$^{-1}$ (typical of young galactic 
clusters) would disperse cluster members over relatively short timescales, $\le$10 Myr.

In addition to the H$\alpha$ emitters concentrated within IC\,1274, there is a dispersed population
that spans the extent of L227. Approximately 8\arcmin\ ($\sim$4.2 pc) southeast of 
IC\,1274, some 14 H$\alpha$ emitters are found near IC\,1275. Most are along the periphery
of L227, but several are near HD 166107, the bright B2 V star at the center of IC\,1275 (Fig. 5).
To the northwest of IC\,1274, a number of H$\alpha$ emitters and X-ray 
sources are distributed along the dark nebulosity leading up to IC\,4684.
This sparse population of T Tauri stars appears to have formed in relative isolation compared
with those within IC\,1274. It is also possible, however, that these stars are simply older
and have drifted from their formation site within the cluster.

\section{The Young Cluster in IC\,1274}

\subsection{The Color-Magnitude Diagram}
The extinction-corrected $(V-I)_{0}$, $V_{0}$ color-magnitude diagram for the H$\alpha$ emission stars and 
X-ray sources with available photometry is shown in Figure 7. The two-dimensional density function of over 
18,000 sources having photometric uncertainties of $\le$0.1 mag in both $V$-band magnitude and $V-I$ color 
from the LFC imaging is represented by the grayscale histogram. In the absence of classification
spectroscopy, we apply the mean extinction value ($A_{V}=1.21$ mag) derived for the B-type cluster stars
to the H$\alpha$ emitters and X-ray sources. This technique was also used by Walker (1957; 1959) in his
studies of NGC\,6530 and IC\,5146 and more recently by Herbig (1998), Herbig \& Dahm (2002), and Dahm \& Simon (2005)
in IC\,348, IC\,5146, and NGC\,2264, respectively. Also shown in Figure 7 is the 100 Myr isochrone of Siess et al.\ (2000),
placed at a distance of 1.82 kpc, which serves as a zero-age main-sequence line. The 0.5, 1.0, and 5.0 Myr
isochrones and the 0.3, 0.5, and 1.0 $M_\sun$ evolutionary tracks of Siess et al.\ (2000), placed in the
observational plane using the main-sequence colors of Kenyon \& Hartmann (1995), are also depicted.
It is apparent that the H$\alpha$ emitters and X-ray sources do not fall along a well-defined pre-main 
sequence, but are instead distributed between the $\sim$0.5 and 5 Myr isochrones. While this may in part
be due to an intrinsic age dispersion, other factors including binarity, variability, and observational
error all contribute to the observed spread. 

\subsection{The Ages and Masses of the H$\alpha$ Emitters and X-ray Sources}
The grid of solar metallicity pre-main-sequence models of Siess et al.\ (2000) was used to derive physical properties for 
the optically detected H$\alpha$ emitters and X-ray sources, assuming the adopted cluster distance and
mean extinction. Uncertainties involved in the
use of pre-main sequence evolutionary tracks and isochrones fall into two broad categories:
the underlying physics used in modeling stellar evolution from the birthline to the zero-age main 
sequence, and the error induced when transforming between theoretical and observational 
planes. Each pre-main sequence model treats convection, opacity, radiative transfer, accretion, and rotation
differently, leading to differences in predicted evolutionary paths for a given stellar mass. Initial
conditions establishing the birthline are also not well understood. Transforming between theoretical
and observational planes is typically achieved by fitting main sequence colors and bolometric corrections
as a function of effective temperature. The use of main sequence colors, however, is questionable
given the enhanced chromospheric activity, accretion, and variability characteristic of 
pre-main sequence stars. The intrinsic colors of such sources, however, are not well established.

Table 5 summarizes the predicted attributes for the H$\alpha$ emitters, including spectral type,
effective temperature ($T_{\rm eff}$), mass, luminosity, radius, and age. Table 6 presents identical data for 
the X-ray sources of Braje et al.\ (2002), excluding the four early-type stars and the H$\alpha$ emitters
included in Table 5. 

The resulting age distribution for the H$\alpha$ emitters and X-ray sources is shown in Figure 8a by the histogram. 
The median of the age distribution is 0.8 Myr, but no corrections have been applied for possible binarity among 
the H$\alpha$ emission and X-ray population of IC\,1274 prior to calculating individual ages and masses. This 
estimated age is sensitive to the adopted distance and would increase by a factor of $\sim$3 if IC\,1274
were 0.5 kpc closer to the Sun.

We show the distribution of masses derived using the Siess et al.\ (2000) models for the H$\alpha$
and X-ray selected sample of stars in IC\,1274 in Figure 8b. The distribution peaks near 0.4 $M_\sun$
before turning sharply over toward lower masses due to survey incompleteness. 
The masses predicted by the pre-main-sequence models of Siess et al.\ (2000) are in reasonable agreement
with the dynamical masses of main-sequence binaries in the mass range from 1.2 to 2.0 $M_\sun$ (Hillenbrand
\& White 2004). In the range from 0.1 to 0.5 $M_\sun$, however, these models predict masses that are 
systematically lower than the dynamically derived values by 5--20\% (Hillenbrand \& White 2004). 
Summing the derived mass values for the detected H$\alpha$ emitters and X-ray sources, we are 
able to account for $\sim$40 $M_\sun$ of the total cluster mass. Assuming an uncertainty of 
$\sim$20\%, we can at best account for $\sim$48 $M_\sun$ of the cluster population. Adding
to this the estimated masses of the four B-type stars, $\sim$41 $M_\sun$, the observed cluster mass
is $\sim$90 $M_\sun$.

The total mass of a cluster can be
approximated by integrating the initial mass function (IMF) in the power-law region of the relation that extends 
from the highest-mass stars to $\sim$0.3 $M_\sun$. This lower mass limit coincides with the least massive members 
of IC\,1274 detected by the H$\alpha$ emission survey. Assuming the Salpeter (1955) IMF power-law slope of 1.35, 
Elmegreen (2000) finds the solution of this integral to be approximated by

$M_{\rm cl} \sim 3\times10^{3} (\frac{M_{\rm max}}{100 M_{\odot}})^{1.35} M_{\odot}$ ,

\noindent where $M_{\rm cl}$ is the total cluster mass and $M_{\rm max}$ is the mass of the most massive star in the cluster 
in units of solar mass. For IC\,1274, this would be the B0 V star HD 166033. Using the average mass of a B0 V star 
from Drilling \& Landolt (2000), 17.5 $M_\sun$, we estimate the total mass of the IC\,1274 cluster to be 
$\sim$285 $M_\sun$. This does not include the flattened, low-mass portion of the IMF below 0.3 $M_\sun$, which 
remains unexplored by the present survey. The implication is that over two-thirds of the cluster's stellar mass content 
remains unaccounted for by the H$\alpha$ emission stars, X-ray sources, and known early-type stars. The undetected 
cluster members either lack strong H$\alpha$ and X-ray emission, are heavily obscured, or remain embedded within 
L227. Deep (100 ks) X-ray imaging of the cluster would likely reveal more candidate cluster members.
Intermediate-mass ($\sim$1.5--3.5 $M_\sun$) stars generally exhibit a large dispersion in fractional X-ray luminosity 
(Feigelson et al.\ 2003; Dahm et al.\ 2007), but should be detectable in a sufficiently deep X-ray image of the cluster. 
Such A-, F-, and G-type stars are notably absent from the H$\alpha$ and X-ray-selected stars. 

\subsection{The Near-Infrared Color-Color Diagram}
Near-infrared $JHK_{S}$ photometry was obtained from the 2MASS Point Source Catalog for 
all sources within a 10\arcmin\ radius of the IC\,1274 cluster center defined by HD 166033. The resulting $J-H$, $H-K_{S}$ 
color-color diagram is shown in Figure 9 for all sources with $J-H$ and $H-K_{S}$ photometric uncertainties 
of $<$0.1 mag. The solid curves are the intrinsic colors for main-sequence stars and giants from Tokunaga (2000),
transformed to the 2MASS filter system using the relations of Carpenter (2000). The dashed parallel lines
are the reddening boundaries for the giant branch and main-sequence loci, derived using the average interstellar
extinction curve of He et al.\ (1995). Significant reddening is apparent for a large population of infrared sources,
presumably distant background stars and galaxies. The majority of the H$\alpha$ emitters and X-ray sources, however, 
suffer little reddening but do exhibit substantial $H-K_{S}$ excess, indicative of dust emission from hot inner 
disks. There appears to be a larger population of infrared excess sources that are
not accounted for by either the H$\alpha$ or the X-ray surveys of the region. Many of these sources lie to 
the right of the reddening line for main-sequence stars and are possible members of IC\,1274 or embedded
Class I protostars within L227.

\section{High-Dispersion Spectroscopy of the Early-type Stars}

The absorption lines of the four B stars are very broad (HD 166033 is the narrowest), so blending
is severe, but all reasonably unblended lines between 3600 and 7200 \AA\ have been measured,
with results given in Table 7. The range in velocity measured for HD 166033 much exceeds the uncertainty
of the individual measurements, so we suspect it to be a spectroscopic binary. The velocity measured
for the sharp \ion{Na}{1} lines is essentially the same for all four B stars, ${-}5.3 \pm 0.2$ km s$^{-1}$,
and is shared by the CH$^{+}$ and \ion{K}{1} interstellar lines in the same spectra. We conclude they
are produced in the foreground.

This can be compared with the mean heliocentric CO velocities in the region determined by Oka et al.\ 
(1999; $-$3.46 km s$^{-1}$), Yamaguchi et al.\ (1999; $-$3.05 km s$^{-1}$), and Fich \& Blitz (1984; $-$2.95 km s$^{-1}$)
that agree that the velocity of L227 is about $-$3.0 km s$^{-1}$. The difference may be because
the CO is concentrated around the periphery of the cavity, while the \ion{Na}{1} cloud is projected
against the B stars in the middle. The velocities of the B stars themselves (in Table 7) are roughly
10 km s$^{-1}$ more positive. Throughout this paper, we have tacitly assumed that the early-type stars and H$\alpha$ emission
populations are essentially contemporary, but to account for the radial velocity discrepancy between
the early-type stars and molecular cloud, we must 
acknowledge the possibility that the B stars formed elsewhere in the foreground and have now moved near
the edge of L227 and are in the process of creating the cavity and exposing the H$\alpha$ emission population. 

\section{The Emission Nebula IC\,1274}

The cavity of IC\,1274 is faintly luminous, but given the spectral types of the illuminating
stars (B0--B5), it is not obvious whether the nebula is likely to be an \ion{H}{2} region, because 
the conventional dividing line between stars that illuminate emission and reflection nebulae 
is about B1. The B0 V star HD 166033 is bolometrically the most luminous of the early-type 
stars, and given its central position within IC\,1274, we believe it to be largely responsible
for the cavity's excavation.

To examine whether the cavity is the remnant of an expanding shell of gas, HIRES spectra were
obtained at three points across the face of IC\,1274 (labeled T1, T2, and T3 in Fig.\ 3). Spectra
of the background at the positions of the B stars have also been recovered. Emission lines
characteristic of an ionized nebula are present: H$\alpha$ is strong and appears as a smooth
structureless Gaussian with a full width at half-maximum (FWHM) of 22 to 24 km s$^{-1}$ with
peak heliocentric radial velocities ranging from $-$5 to $-$7 km s$^{-1}$. The intensity ratio of the
[S\ II] lines at $\lambda\lambda$6717, 6730 is an indicator of electron density. Following the
formalism of Keenan et al.\ (1996), we derive an electron density of about 100 cm$^{-3}$, demonstrating
that the cavity contains a very weakly ionized emission nebula, but there may also be scattered light
from the other B-type stars as well.

If the cavity were an expanding shell, the radial velocities of the narrow [S\ II] and [N\ II]
emission lines should vary systematically with respect to the distance from the center of
the nebula (i.e., HD 166033). At the assumed distance, and a time interval of 1 Myr, the expansion
velocity would be $\sim$1--2 km s$^{-1}$. The emission lines, however, remain single, at $-$3
to $-$7 km s$^{-1}$, with the exception of those at HD 166033, where they are double and have a 
separation of 8 to 10 km s$^{-1}$. Provided in Table 8 are the rectangular coordinates in arcsec
of T1--T3 and the B-type stars relative to HD 166033, the radial distance from HD 166033 in arcsec,
the measured heliocentric radial velocity of [S\ II] emission at these points, and the intensity
ratio of 6717/6730. These results place a firm restraint on any speculation that the diameter 
of the shell is correlated with the dynamic age of IC\,1274.

\section{Discussion}

Our interpretation of the structure of IC\,1274 is as follows: The early-type stars in the cluster 
formed recently, and have since ionized and dissociated the H$_{2}$ and CO on the near side of the 
molecular cloud, forming the cavity that is now observed. The faintly luminous spherical shell is
outlined particularly on the east and southeast by what appears to be a pileup of material
driven there, directly or indirectly, by the B-type stars near the center. The western edge of the nebula 
is not so clearly defined. There is a striking concentration of H$\alpha$ emitters in the southeastern 
half of the cavity. An explanation might be that the molecular gas that originally filled the cavity 
was disrupted when the B stars formed, and material in the eastern half was driven up against what now 
survives as the massive L227 molecular cloud. To the west there was no such resistance, and material flowed 
freely outward in that direction. The increase in density to the east enabled the formation of low-mass 
stars in that volume. If so, the photometric ages of those TTSs ($\sim$1 Myr) would have to be compatible 
with the time required for the cavity to have been formed. An alternate explanation is that the TTSs may
have been present well before the early-type stars formed, and are only now detectable because the 
expanding cavity cleared away the foreground extinction. 

Substantial low-mass star formation has accompanied the appearance of the massive stars and indeed may have 
preceded their formation if the isochronal ages of some individual H$\alpha$ emitters and X-ray sources are 
taken at face value. Limited low-mass star formation is occurring along the periphery of the L227 molecular 
cloud from IC\,4684 to IC\,1275, some $\sim$10 pc in extent. Shown in Figure 10 is a 1$^{\circ}$ square
{\it IRAS} composite image (25, 60, and 100 $\mu$m) centered upon IC\,1274 with the positions of the early-type 
stars, H$\alpha$ emitters, and X-ray sources superimposed. Dust emission peaks within the cluster core, which 
is offset from the dark nebulosity evident on optical images of the field (compare with Figs.\ 1, 2, and 5). Some
$\sim$20\arcmin\ south of IC\,1274, NGC\,6559 exhibits significant far-infrared emission, presumably from
dust heated by the presence of early-type stars that illuminate the emission/reflection nebula. The H$\alpha$
emission stars and X-ray sources effectively trace the areas of warm dust emission.

Oka et al.\ (1999) mapped the L227 molecular cloud in the $J=1-0$ transition of CO and found a mean cloud 
velocity that corresponds to a kinematic distance of $1.7\pm1.0$ kpc. The virial mass of 
the cloud agrees well (factor of 2) with that derived using the total CO luminosity and a standard CO to 
H$_{2}$ mass ratio (Scoville et al.\ 1987). This total mass, 1--$2\times10^{4} M_{\sun}$, is comparable
to that of the molecular clouds in the Taurus-Auriga complex (Mizuno et al.\ 1995). The cloud exhibits a 
sharp, well-defined boundary (see Fig.\ 1 of Oka et al.\ 1999) that parallels the eastern edge of dust 
emission evident in Figure 10. The Very Large Array (VLA) 21 cm continuum observations of the region by Braje et al.\ (2002) 
reveal a spherical mass of \ion{H}{1} emission coincident with IC\,1274 (see their Fig.\ 1). The neutral 
hydrogen emission likely originates from the photodissociation of H$_{2}$ by the early-type stars. 
The conically shaped \ion{H}{1} emission region north of IC\,1274 evident in the VLA map has been the subject of 
an intense search to isolate the Galactic $\gamma$-ray source GeV J1809$-$2327.  

The identity of GeV J1809$-$2327 could have some bearing on the early star forming history of L227. If a young 
pulsar is responsible for the time-variable $\gamma$-ray emission as hypothesized by Oka et al.\ (1999) 
and Braje et al.\ (2002), its progenitor supernova could have triggered the active star formation that 
is occurring throughout L227 and the Simeis 188 nebula complex. The low spatial resolution ($\sim$3\arcmin) 
{\it ASCA} Gas Imaging Spectrometer (GIS) survey of Oka et al.\ (1999) identified two prominent X-ray sources separated by $\sim$6\arcmin. 
The southern source is pointlike, but appears connected with the IC\,1274 cluster. X-ray emission for this 
source peaks between 0.5 and 2.2 keV, primarily due to the response of the ACIS detectors. The northern point 
source is associated with diffuse X-ray synchrotron emission and was considered the likely candidate for the 
$\gamma$-ray source. This X-ray source exhibits a significant hard component from 2.2 to 10 keV that was not 
observed for the southern source. Within the synchrotron nebula, Oka et al.\ (1999) speculate that a pulsar 
is producing high-energy electrons (TeV-level) that are impacting the molecular gas of L227, some 30\arcsec\
or $\sim$0.3 pc distant, rapidly decelerating, and emitting the observed $\gamma$-ray flux. 

The higher-resolution (0\farcs3) {\it Chandra} ACIS and VLA 21 cm observations of the region by 
Braje et al.\ (2002) revealed an X-ray point source near the {\it ASCA} GIS detection that is connected to a 
nonthermal X-ray and radio emission nebula. Their X-ray source 5 lacks an optical counterpart on the 
Palomar Sky Survey red image and appears connected to the diffuse radio nebula by a 30\arcsec\ long jet of X-ray
emission. This source exhibits a composite X-ray spectrum with a hard component as well as an underlying thermal 
emission spectrum. Braje et al.\ (2002) propose that source 5 is a young pulsar (age $\sim$10$^{4}$--10$^{6}$ yr),
and that its wind and nebula are responsible for the observed $\gamma$-ray emission.

The LFC observations presented here include the location of source 5. To identify an optical counterpart, 
we first registered a dozen X-ray sources having well-defined, unsaturated optical counterparts and determined
a median relative position error between the ACIS and LFC astrometry of $\sim$0.9$\pm$0.4\arcsec.
The offset for source 5 specifically with its proposed optical counterpart is $\sim$1.2\arcsec. The
source is faint, $V\sim21.5$, and if at the adopted distance of IC\,1274, it would lie near the 0.5 Myr 
isochrone of Siess et al.\ (2000), several magnitudes above the zero-age main sequence with a mass of
$\sim$0.25 $M_\sun$.

Shown in Figure 11 is a three-color composite ($VRI$) image of source 5 taken from the LFC imaging.
The synchrotron emission nebula extends to the northwest, toward the luminous star causing the halo 
evident in the upper right corner of Fig. 11. The LFC imaging reveals a peculiar red flaring that extends 
several arcseconds south of source 5 that may be a confused background object. Given the densely populated 
field, it is certainly possible that source 5 and the proposed optical counterpart are the result of a 
random superposition. Considering that the $\gamma$-ray line of sight extends far beyond L227, and the
lack of any persuasive structural connection between the two, we conclude that the proposed causal 
connection between the $\gamma$-ray source and star formation activity in IC\,1274 and L227 has yet
to be demonstrated.

\section{Summary and Conclusions}

IC\,1274 is a faintly luminous emission nebula enveloping a spherical cavity excavated out of the
L227 molecular cloud very probably by the B0 V star HD 166033. This single-line spectroscopic binary hosts a small
clustering of fainter stars within a distance of $\sim$20\arcsec\ or projected separations of $\le$0.18 pc.
From the intensity ratio of the forbidden [S\ II] lines at $\lambda\lambda$6717, 6730, we derive an 
electron density of $\sim$100 cm$^{-1}$, establishing IC\,1274 as a weakly ionized emission nebula,
but scattered light from the other B-type stars must also be present.

Star formation, as evidenced by the presence of numerous H$\alpha$ emission stars and X-ray sources, 
presumably members of a low-mass, pre-main-sequence population, is occurring within and around IC\,1274
and the L227 molecular cloud. Over $\sim$80 stars with H$\alpha$ in emission and brighter than $V\sim21$
have been identified in the region as well as 21 X-ray sources (Braje et al.\ 2002), many of which are 
coincident with the early-type stars and H$\alpha$ emission stars in IC\,1274. Correcting for interstellar
extinction and assuming a distance of 1.82 kpc (inferred from $BV$ photometry of the four B-type stars),
these H$\alpha$ emitters and X-ray sources lie $\sim$1--3 mag above the zero-age main sequence. A median age
of $\sim$1 Myr is derived using the evolutionary models of Siess et al.\ (2000), although a significant dispersion 
is present.

The H$\alpha$ emission stars and X-ray sources are strongly concentrated around the massive B-type stars 
in IC\,1274, and are preferentially positioned in the southeastern half of the nebula. The dust and molecular
gas that once filled the cavity within IC\,1274 may have been driven up against the massive L227 molecular 
cloud to the east when the B stars formed. The increase in density may have enabled the formation of
the low-mass stars in that volume. To the west, the outflowing gas met no resistance and flowed freely into
the interstellar medium. Some 14 H$\alpha$ emission stars are identified along the periphery of the L227
cloud near IC\,1275 $\sim$10\arcmin\ or $\sim$5 pc to the south. Another handful of emitters is found along
the lanes of dark nebulosity leading up to the IC\,4684 reflection nebula.

Notably absent from IC\,1274 are bright H$\alpha$ emission stars that could be intermediate-mass Herbig AeBe 
or classical Be stars. Given the relative youth of the cluster, the presence of Herbig AeBe stars would be 
expected. We identify one candidate classical Be star (LkH$\alpha$ 127) lying near IC\,4684. This source 
exhibits strong H$\alpha$ emission, but its near-infrared colors are consistent with photospheric emission.

Assuming that the early-type stars and detected H$\alpha$ emitters and X-ray sources form a real cluster, we conclude
that only one-third of the cluster mass is accounted for if a Salpeter-like IMF is adopted. Deeper 
high-resolution X-ray imaging of IC\,1274 is needed to identify the bulk of the cluster stars, 
particularly the intermediate-mass members that eluded detection by the H$\alpha$ emission survey 
and the shallow {\it Chandra} survey of Braje et al.\ (2002). In press it was brought to the attention
of the authors that a deeper (30 ks) {\it Chandra} ACIS-I observation was recently made of this region 
(ObsID 12546). When available, this image should detect a substantial number of additional
cluster members.

The faint ($V\sim21.5$) optical counterpart of X-ray source 5 identified in the LFC imaging exhibits colors 
that are consistent with a low-mass, pre-main-sequence star if at the adopted distance of IC\,1274. Given
the density of the field star population in this region, however, it is not unreasonable to assume that the 
X-ray source and optically detected star are randomly superposed. We conclude that the linkage between the 
$\gamma$-ray source GeV J1809-2327, X-ray source 5, and star formation activity within L227 is derived
from the concatenation of several hypotheses that requires additional evidence to support. The deeper
{\it Chandra} ACIS-I observation of the region may resolve some of these issues.

\acknowledgments
We have made use of the Digitized Sky Surveys, which were produced at the Space Telescope Science Institute 
under U.S. Government grant NAG W-2166, the SIMBAD database operated at CDS, Strasbourg, France, and the 
Two Micron All Sky Survey (2MASS), a joint project of the University of Massachusetts and the Infrared 
Processing and Analysis Center (IPAC)/California Institute of Technology, funded by NASA and the National 
Science Foundation. GHH's participation in this investigation was partially supported by the National
Science Foundation under grant AST 07-02941. We are very grateful to Bo Reipurth for his invaluable assistance in obtaining the 2010
high-dispersion spectra presented here. We also gratefully acknowledge the Canada-France-Hawaii Telescope 
\& Coelum / J.-C. Cuillandre \& G. Anselmi for use of the true-color image of IC\,1274 and L227 reproduced
in Fig.\ 1. We express gratitude to Brian Patten for his assistance in obtaining the 1990 wide-field grism 
observations. SED would also like to thank Roy Gal for his detailed summary of the LFC reduction pipeline,
and BPB gratefully acknowledges Lucas Cieza for assistance with the 2009 WFGS2 observations. Finally we
wish to thank our referee, Leisa Townsley, for her detailed comments and suggestions that significantly 
improved this manuscript.

Facilities: \facility{Keck 1}, \facility{Hale}, \facility {UH 2.2 m}

\singlespace
\clearpage
\begin{figure}
\epsscale{1.0}
\plotone{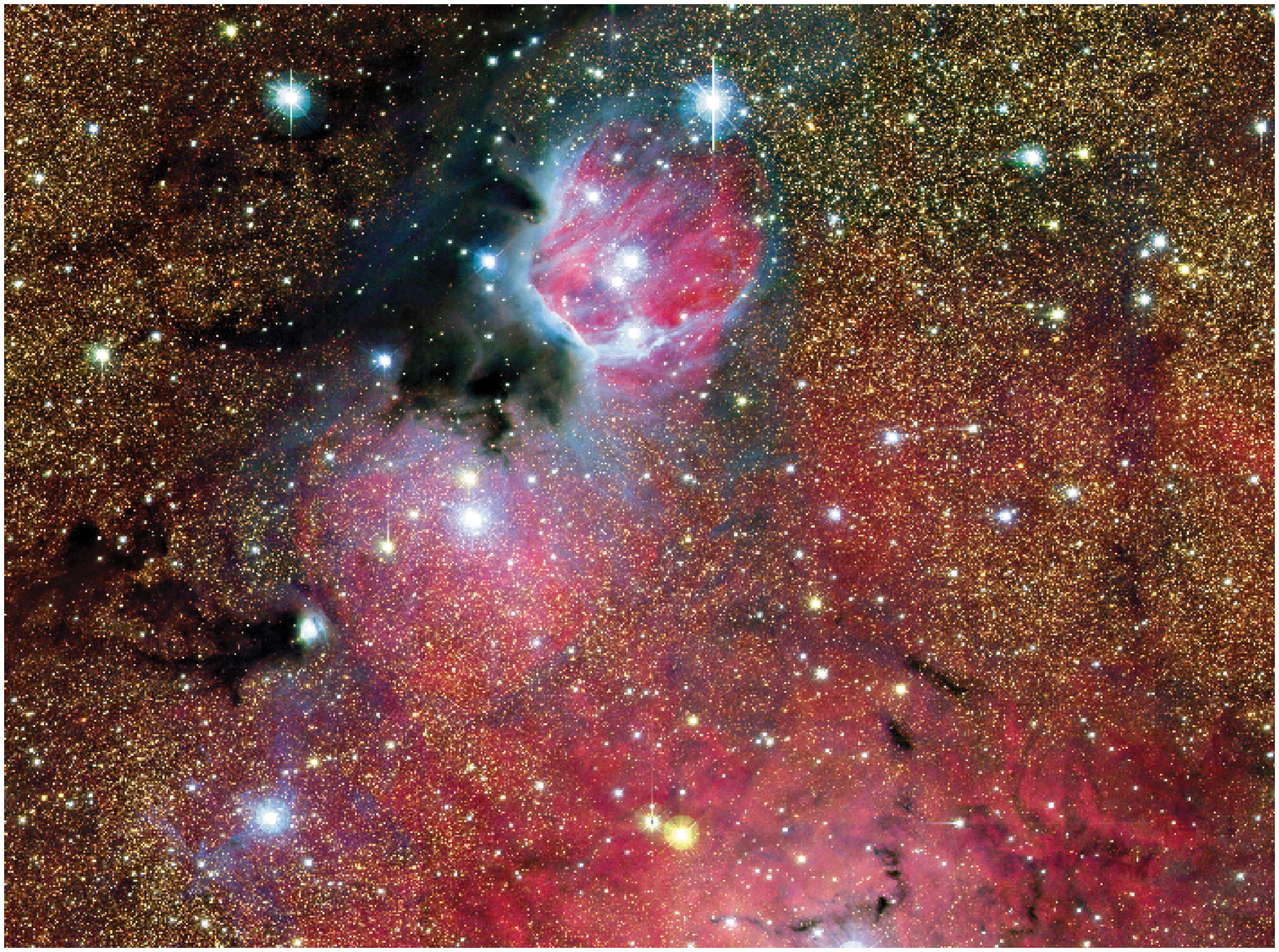}
\caption[f1.eps]{A true-color ($BVR$) image of IC\,1274 and L227 obtained by
Canada-France-Hawaii Telescope \& Coelum / J.-C. Cuillandre \& G. Anselmi using the CFH12K wide-field
imaging camera and reproduced here with permission. North is at top and east to the left.
The approximate center of the field of view is: $\alpha_{J2000}=18^{\rm h}09^{\rm m}39^{\rm s}$, $\delta=-23{\arcdeg}46\arcmin40\arcsec$.
IC\,1274 is the luminous spherical cavity $\sim$5\arcmin\ in diameter that appears to have been carved out of
L227, presumably by the early-type stars within. The illuminated rim of L227 is
clearly defined along the eastern periphery of IC\,1274. To the west it is evident that the opacity
of remnant molecular material is significantly reduced.
\label{f1}}
\end{figure}
\clearpage

\clearpage
\begin{figure}
\epsscale{0.8}
\plotone{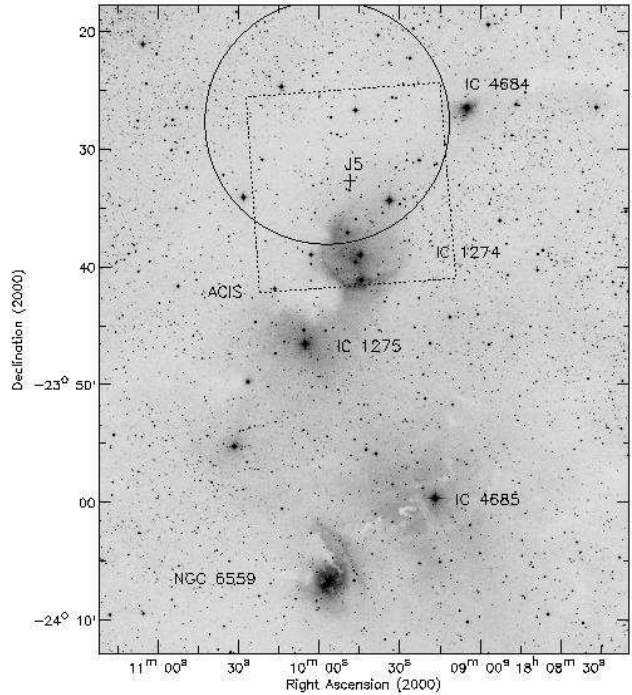}
\caption[f2.eps]{The Simeis 188 complex of emission and reflection nebulae, and molecular clouds. This 
field is centered near $\alpha_{J2000}=18^{\rm h}09^{\rm m}45^{\rm s}, \delta=-23^{\circ}45\arcmin$ and is taken from the 
blue Digitized Sky Survey image of the region. The orientation is as in Fig.\ 1; the field of view 
is $45\arcmin\times56\arcmin$. Prominent components of Simeis 188 are identified, including IC\,4684,
IC\,1274, IC\,1275, IC\,4685, and NGC\,6559. The cross-marked J5 is the position of the {\it Chandra}
X-ray source 5 ($180950.2-233223$) of Braje et al.\ (2002). The large circle represents the 95\% 
confidence circle for the $\gamma$-ray point source GeV J1809$-$2327. The ACIS field of view for 
the shallow {\it Chandra} observation of the region is represented by the dotted square.
\label{f2}}
\end{figure}
\clearpage

\clearpage
\begin{figure}
\epsscale{1.0}
\plotone{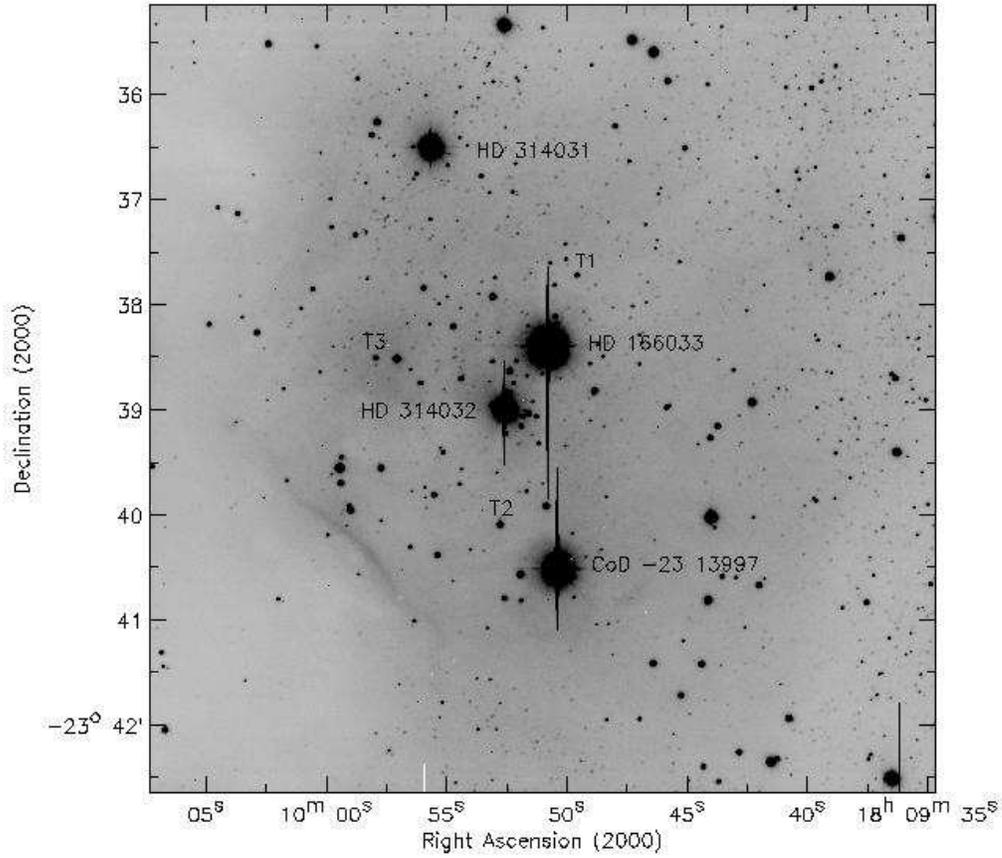}
\caption[f3.eps]{The center of IC\,1274 taken from an $R$-band image obtained at the f/10 focus of the 
UH 2.2 m telescope. The field is approximately 7\farcm5 on a side and the orientation is as in Fig.\ 1.
The four early-type stars in the emission nebula are labeled, as are the three points where HIRES spectra
were obtained of the nebula (labeled T1, T2, and T3).
\label{f3}}
\end{figure}
\clearpage

\clearpage
\begin{figure}
\epsscale{1}
\plotone{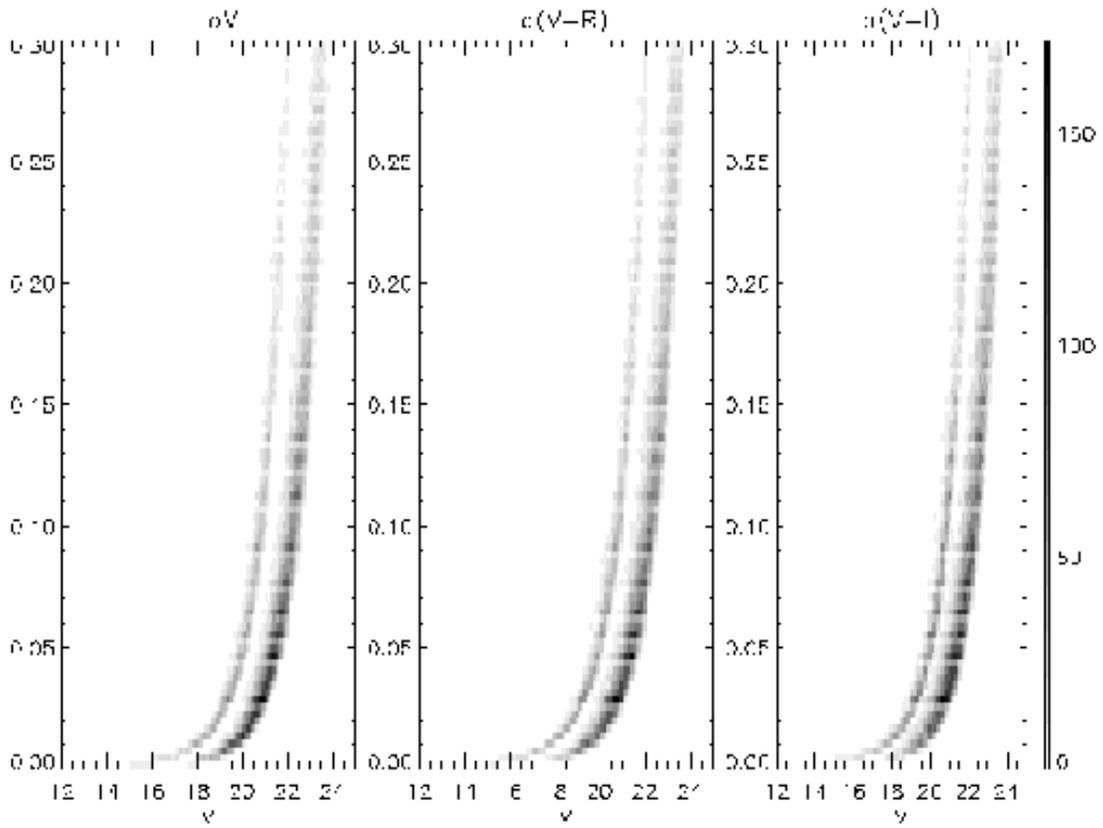}
\caption[f4.eps]{The density function histogram of internal photometric errors of $V$-band magnitude, 
($V-R$), and ($V-I$) colors as a function of $V$ for short and long exposures obtained using the Large 
Format Camera on the Hale 200 inch (5 m) telescope. Photometric errors for most sources brighter than
$V\sim$22 are $\le$0.1 mag.
\label{f4}}
\end{figure}
\clearpage

\clearpage
\begin{figure}
\epsscale{0.9}
\plotone{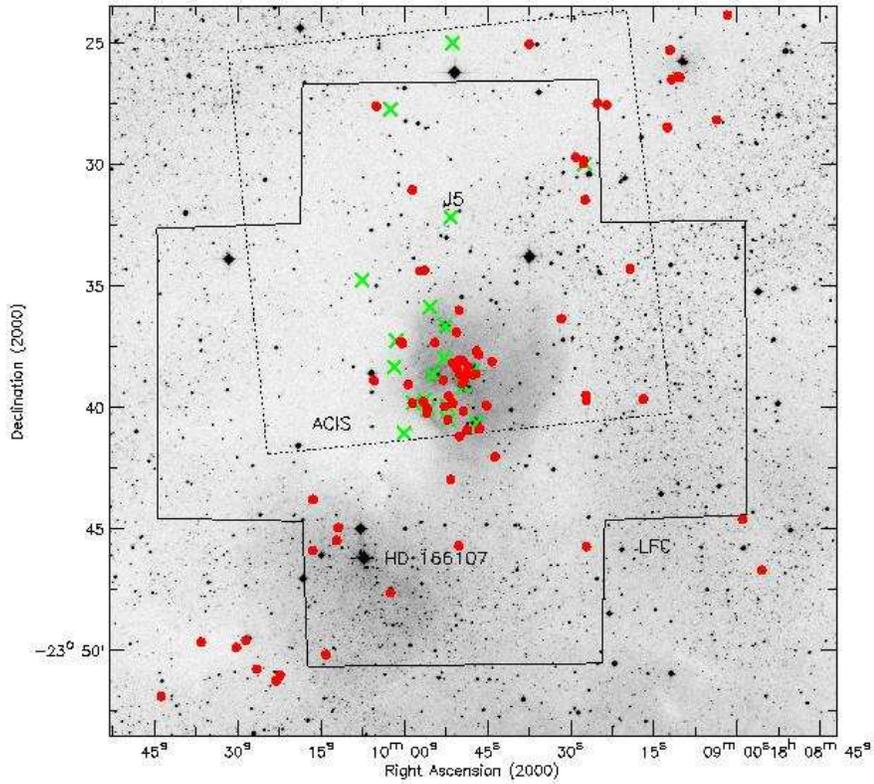}
\caption[f5.eps]{A $30\arcmin\times30\arcmin$ red image of IC\,1274 obtained from the Digitized 
Sky Survey with the positions of the H$\alpha$ emission stars and X-ray sources of Braje et al.\ (2002) 
marked by red circles and green crosses, respectively. Also shown are the outlines of the LFC and
{\it Chandra} ACIS fields of view, which encompass much of the region. Most of the H$\alpha$ emitters are concentrated
within IC\,1274, and a striking number are located in the southeastern half of the nebula. We suggest
that the molecular gas that originally filled the cavity was disrupted by the early-type stars. 
Material in the eastern half was driven up against what now survives of the massive L227 molecular cloud.
\label{f5}}
\end{figure}
\clearpage

\clearpage
\begin{figure}
\plotone{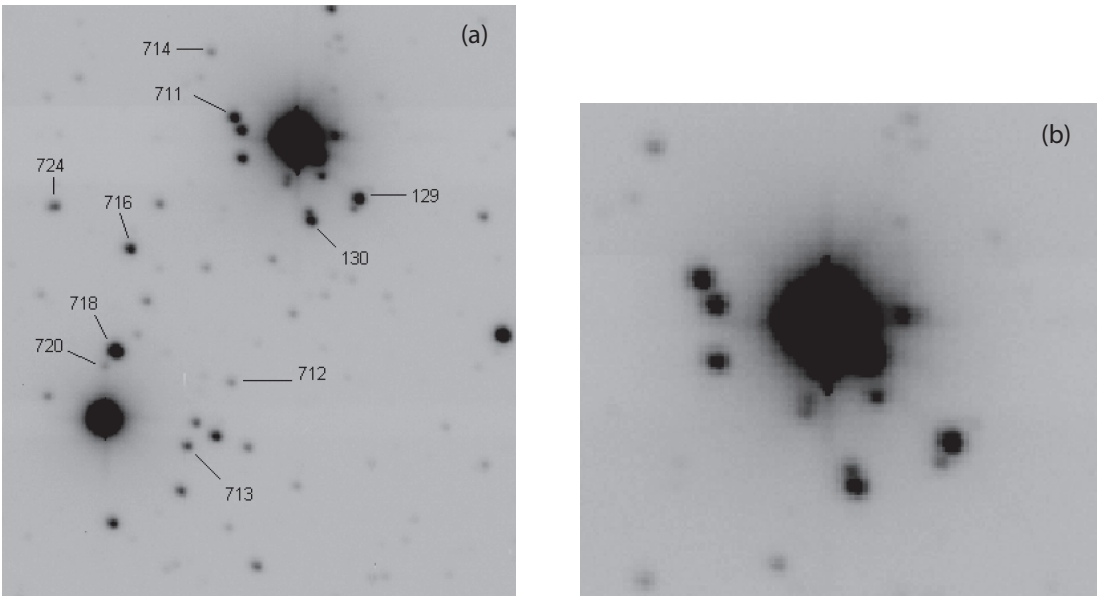}
\caption{(a) $I$-band image obtained at the f/10 focus of the UH 2.2 m- 
telescope of HD 166033 (upper right) and HD 314032 (lower left). The field of view is approximately
$1\farcm1\times1\farcm25$. Several H$\alpha$ emitters in the region are identified by their
IH$\alpha$ numbers. (b) The immediate vicinity of HD 166033 ($\sim$27\arcsec\ square
field of view) revealing a dense clustering of fainter sources not observed around the 
other B-type stars.
\label{f6}}
\end{figure}
\clearpage

\clearpage
\begin{figure}
\epsscale{0.8}
\plotone{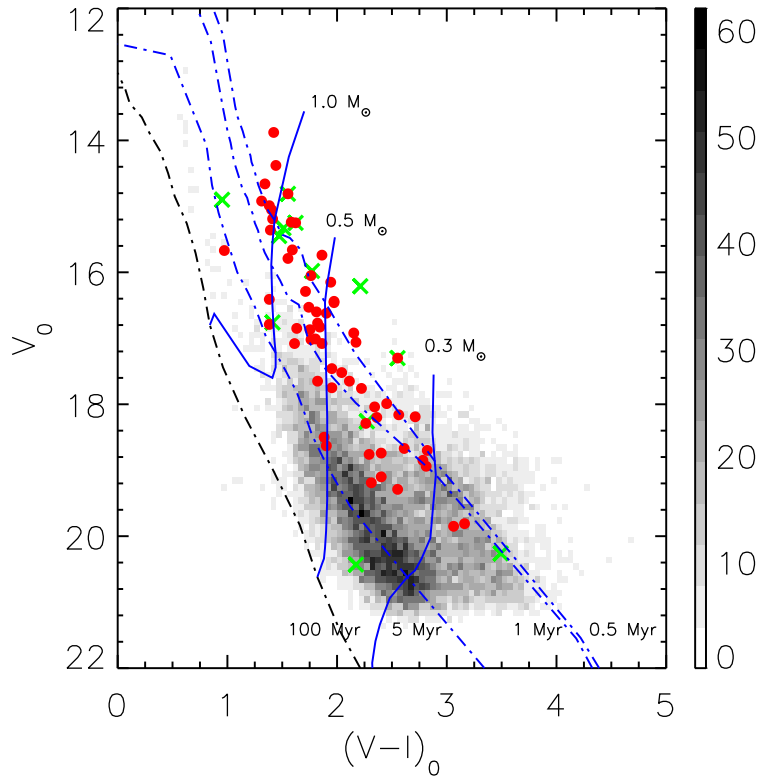}
\caption[f7.eps]{Extinction-corrected $(V-I)_{0}$, $V_{0}$, color-magnitude diagram of the IC\,1274 region.
The mean extinction value derived from photometry of the early-type stars has been applied 
to all H$\alpha$ emission stars and X-ray sources. The two-dimensional density function of over 18,000
sources having photometric uncertainties of $\le$0.1 mag in both $V$-band magnitude and $V-I$ color
from the LFC imaging is represented by the gray-scale histogram. H$\alpha$ emission stars are shown
as solid red points, and X-ray sources from Braje et al.\ (2002) by green crosses. The black dashed 
line is the 100 Myr isochrone of Siess et al.\ (2000) placed at 1.82 kpc. The blue dashed lines are the
0.5, 1.0, and 5.0 Myr isochrones from the same pre-main-sequence models. The solid vertical curves are
0.3, 0.5, and 1.0 $M_\sun$ evolutionary tracks.
\label{f7}}
\end{figure}
\clearpage

\clearpage
\begin{figure}
\plotone{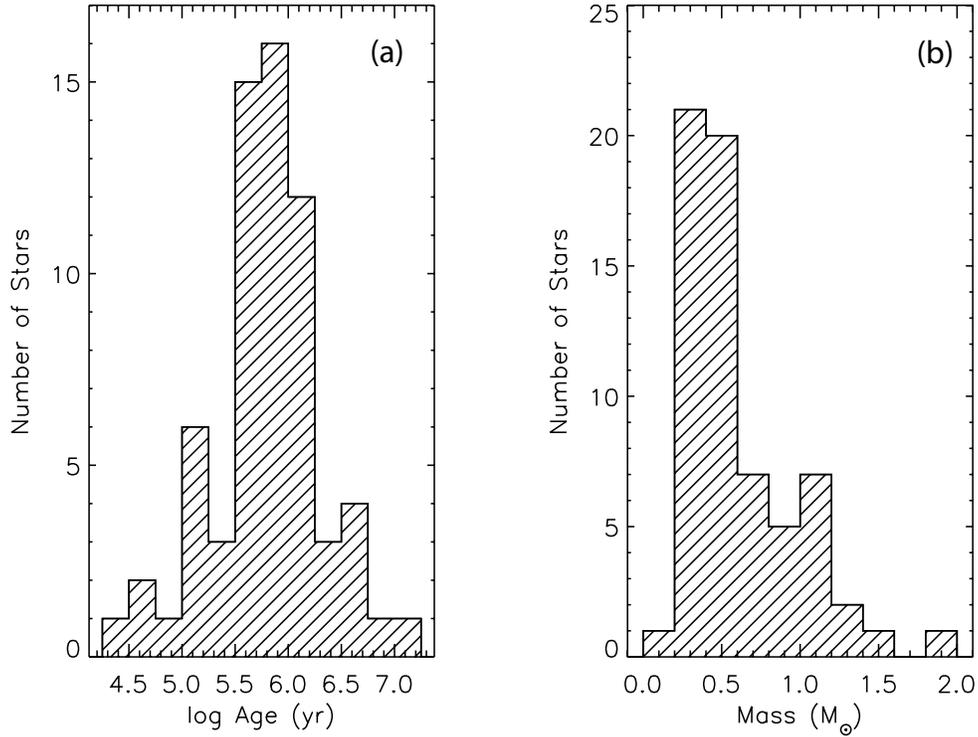}
\caption[f8.eps]{Histograms of (a) age and (b) mass  as predicted by the models of Siess et al.\ (2000)
for the H$\alpha$ emission stars and X-ray sources in the IC\,1274 region with optical photometry available. 
The median age of the distribution is $\sim$0.8 Myr, but no corrections for binarity have been applied. The
mass distribution peaks near 0.4 $M_\sun$ before turning over sharply toward lower masses due to survey
incompleteness.
\label{f8}}
\end{figure}
\clearpage

\clearpage
\begin{figure}
\epsscale{0.85}
\plotone{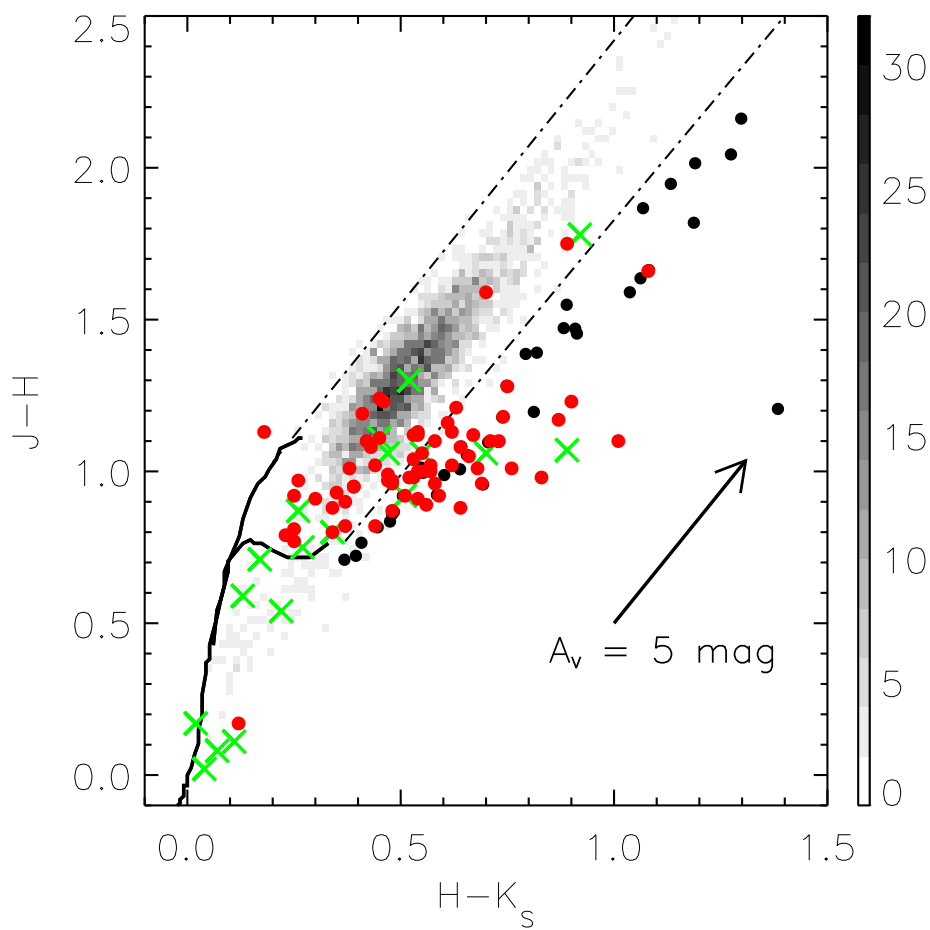}
\caption[f9.eps]{$H-K_{S}$,$J-H$ two-color diagram for all sources in the 2MASS Point Source Catalog within 10\arcmin\ of 
HD 166033 and having photometric uncertainties $<$0.1 mag. Solid curves represent the intrinsic colors of 
main-sequence stars
and giants from Tokunaga (2000) placed on the 2MASS system using the transformations of Carpenter (2000). The dashed parallel 
lines are the reddening boundaries for the giant branch and main-sequence loci, derived using the average interstellar
extinction curve of He et al.\ (1995). Symbols are as in Figure 5 with the addition of solid black circles that represent 
2MASS infrared excess sources that are unaccounted for by either the H$\alpha$ or the X-ray surveys of the region.
\label{f9}}
\end{figure}
\clearpage

\clearpage
\begin{figure}
\epsscale{0.85}
\plotone{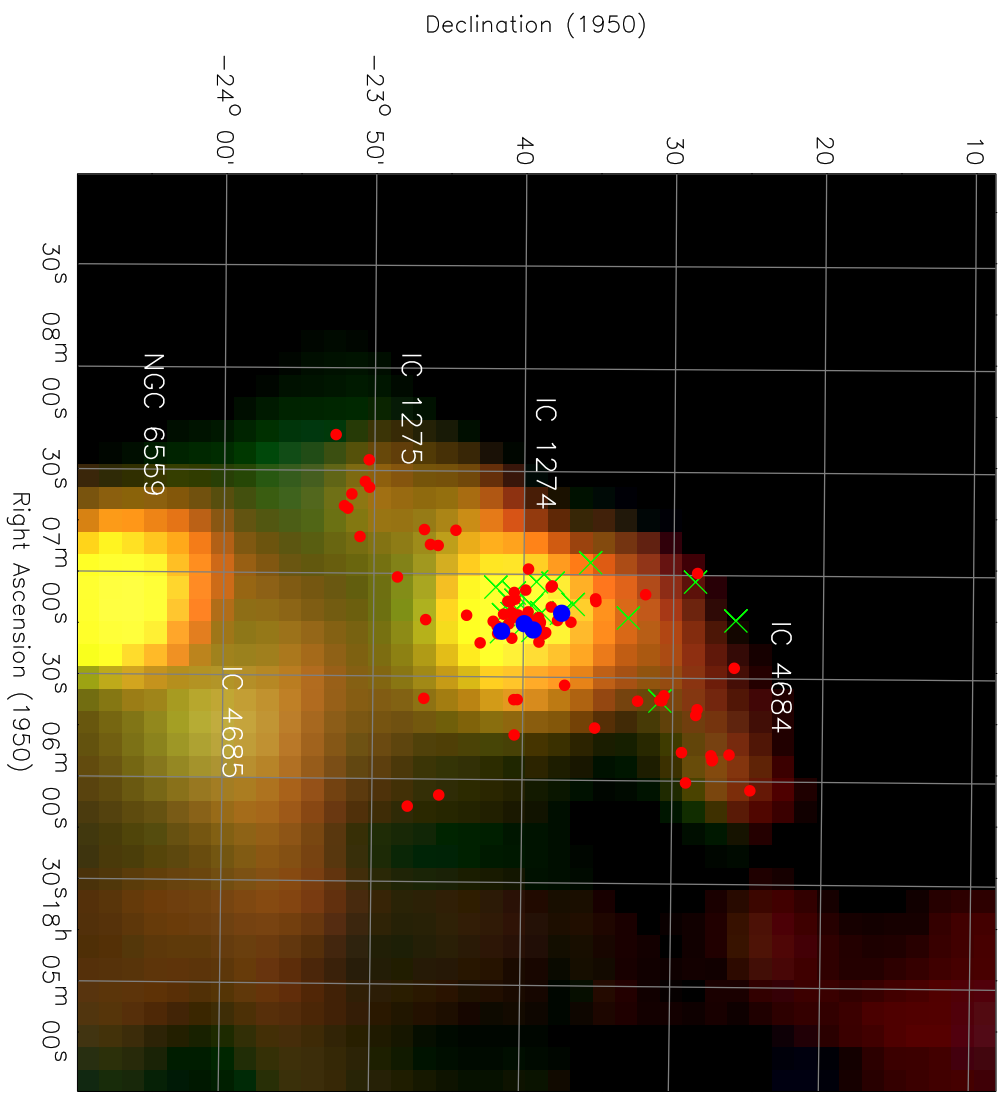}
\caption[f10.eps]{IRAS composite image (25, 60, 100 $\mu$m) of the IC\,1274 region (compare with Figs. 1, 2 and 5). 
B1950 coordinates are retained for convenience when comparing to the IRAS point source catalog. 
The early-type stars, H$\alpha$ emission stars, and X-ray sources are represented by solid blue points, 
solid red points, and green crosses, respectively. Dominant components of the Simeis 188 complex are 
identified. It is noted that the H$\alpha$ emitters and X-ray sources trace out regions of warm dust 
emission. The prominent mass of dust emission south of IC\,1274 is centered upon the emission/reflection
nebula NGC\,6559. The regions south of $-$23$^{\circ}55$\arcmin\ and north of $-$23$^{\circ}20$\arcmin\
were not examined by this investigation.
\label{f10}}
\end{figure}
\clearpage

\clearpage
\begin{figure}
\epsscale{1}
\plotone{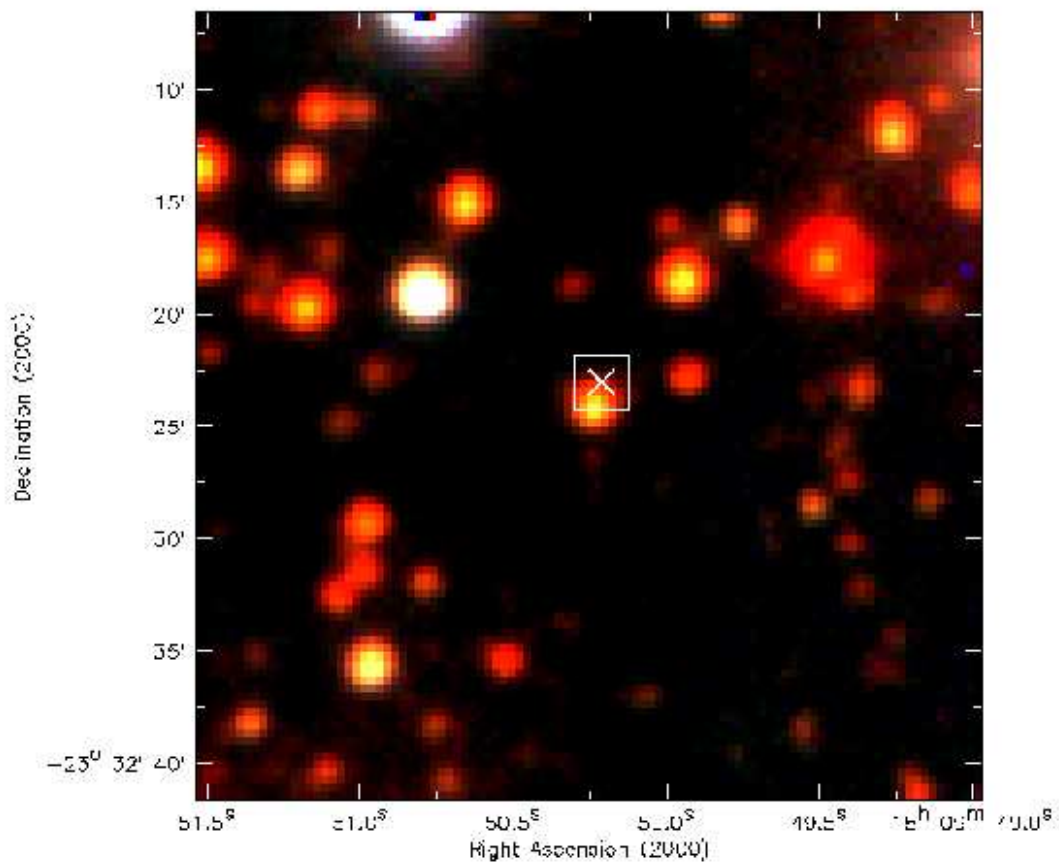}
\caption[f11.eps]{$VRI$ composite image of the region near X-ray source 5 of Braje et al.\ (2002) obtained 
using the LFC on the Hale 200 inch (5 m) telescope. That source is the suspected pulsar responsible for the
$\gamma$-ray emission in the region. The central white cross marks the position of the {\it Chandra} X-ray
point source, and the surrounding box represents the uncertainty in position of the X-ray source relative
to the LFC astrometry. The median position error between the ACIS and LFC astrometry for a dozen X-ray sources
with well-defined optical counterparts is $\sim$0.9\arcsec. There is a faint star ($V$=21.5) very near the
position marked, but given the star's lack of distinctive characteristics, coupled with the density of faint
stars in the area, it is unclear if this particular object should be identified with the {\it Chandra} source.
\label{f11}}
\end{figure}
\clearpage

\begin{deluxetable}{cccccccc}
\tablenum{1}
\tablewidth{0pt}
\tablecaption{Early-Type Stars in IC\,1274}
\tablehead{
\colhead{Star}  & \colhead{CoD}  & \colhead{$V$\tablenotemark{a}}  & \colhead{$B-V$\tablenotemark{a}} & \colhead{Sp Type\tablenotemark{b}} & \colhead{$A_{V}$} & \colhead{Distance} &\colhead{J source\tablenotemark{c}}\\
&&&&&& \colhead{(kpc)}
}
\startdata
...                         &  $-$23$^{\circ}$13997 &  9.16  &  +0.07 &    B1 V   &  1.04    &  1.84  & 12 \\
HD 166033\tablenotemark{d}  &  $-$23$^{\circ}$13998 &  8.59  &  +0.07 &    B0 V   &  1.15    &  1.94  & \phn1  \\
HD 314032                   &  $-$23$^{\circ}$13999 &  9.96  &  +0.22 &    B5 V   &  1.21    &  0.98  & 13 \\
HD 314031\tablenotemark{e}  &  $-$23$^{\circ}$14002 &  9.89  &  +0.20 &    B0.5 V &  1.49    &  2.51  & \phn3 \\
\enddata
\tablenotetext{a}{Optical photometry from Herbst et al.\ (1982).}
\tablenotetext{b}{Spectral types from Herbst et al.\ (1982) and Crampton \& Fischer (1974).}
\tablenotetext{c}{Source identifier from Braje et al.\ (2002).}
\tablenotetext{d}{Some confusion arose because the identification of these stars on the HDE charts
was not compatible with the original HD listing. Herbst et al.\ (1982) followed the HDE numbering,
where our HD 166033 was identified as HD 166079, and our CoD $-$23$^{\circ}$13997 was called HD 166033.}
\tablenotetext{e}{This star was once believed to be a white dwarf (WD 1806$-$23) on the basis of color
and a large proper motion. The proper motion was shown to be in error by Grasdalen (1975), and there
is now no reason to believe that the star is not at about the distance of IC\,1274.}
\end{deluxetable}

\begin{deluxetable}{lllc}
\tablenum{2}
\tablewidth{0pt}
\tablecaption{Distance Estimates for Sh 2-32}
\tablehead{
 \colhead{Reference}  & \colhead{Method}  & \colhead{Technique} & \colhead{Distance} \\
 &&& \colhead{(kpc)}
}
\startdata
This paper              &  Photometric       &  B stars $BV$          & $1.82\pm0.3$ \\
Georgelin et al.\ (1973)\tablenotemark{a} &  Kinematic         &  H$\alpha$ velocities  & $1.88\pm1.03$ \\
Georgelin et al.\ (1973)&  Photometric       &  B stars $UBV$, H$\beta$         & 2.4           \\
Vogt \& Moffat (1975)\tablenotemark{b}   &  Photometric       &  B stars $UBV$         & 2.2           \\
Fich \& Blitz (1984)   &  Kinematic         &  CO velocities         & $1.80\pm0.60$ \\
Oka et al.\ (1999)     &  Kinematic         &  CO velocities         & $1.7\pm1.0 $  \\
\enddata
\tablenotetext{a}{H$\alpha$ interferometric kinematic distance for the \ion{H}{2} region.}
\tablenotetext{b}{Derived a mean extinction of $A_{V}\sim$1.17 mag.}
\end{deluxetable}

\begin{deluxetable}{ccccccccccccc}
\tabletypesize{\tiny}
\rotate
\tablenum{3}
\tablewidth{0pt}
\tablecaption{H$\alpha$ Emission Stars in the IC\,1274 Region}
\tablehead{
\colhead{IH$\alpha$\tablenotemark{a}}  & \colhead{$\alpha$}  & \colhead{$\delta$}  & \colhead{$V$\tablenotemark{b}} & \colhead{$V-R$\tablenotemark{b}} & \colhead{$V-I$\tablenotemark{b}} & \colhead{$J-H$\tablenotemark{c}} & \colhead{$H-K_{S}$\tablenotemark{c}} & \colhead{$K_{S}$\tablenotemark{c}} & \colhead{W(H$\alpha$) 1990\tablenotemark{d}} & \colhead{W(H$\alpha$) 2003\tablenotemark{d}} & \colhead{W(H$\alpha$) 2009\tablenotemark{d}} & Comments\\
                &   (J2000)           &   (J2000)           &               &                 &                 &                 &                     &                   &   (\AA)  & (\AA) & (\AA)    
}
\startdata
686  &  18 08 55.18  &  $-$23 47 05.4  & ...   & ...   & ...  &  1.19 &  0.41 & 11.64 &  ...      & ...      & $-$105:   & \\
687  &  18 08 58.47  &  $-$23 44 59.7  & ...   & ...   & ...  &  1.02 &  0.62 & 11.73 &  ...      & ...      & $-$13   & \\
688  &  18 09 00.03  &  $-$23 24 15.2  & ...   & ...   & ...  &  1.02 &  0.57 & 10.97 &  ...      & ...      & $-$5    & \\
689  &  18 09 02.26  &  $-$23 28 32.8  & ...   & ...   & ...  &  0.99 &  0.47 & 11.82 &  ...      & ...      & $-$38   & \\
690  &  18 09 08.75  &  $-$23 26 47.3  & ...   & ...   & ...  &  0.96 &  0.58 & 11.31 &  ...      & ...      & $-$7    & \\
691  &  18 09 09.10  &  $-$23 26 45.1  & ...   & ...   & ...  &  0.82 &  0.44 &  9.06 &  ...      & ...      & $-$2      & \\
692  &  18 09 10.18  &  $-$23 26 51.6  & ...   & ...   & ...  &  0.79 &  0.23 & 11.40 &  ...      & ...      & EM        & \\
693  &  18 09 10.43  &  $-$23 25 39.3  & ...   & ...   & ...  &  0.92 &  0.25 & 12.12 &  ...      & ...      & $-$7:     & \\
694  &  18 09 11.09  &  $-$23 28 49.6  & ...   & ...   & ...  &  1.24 &  0.45 & 11.74 &  ...      & ...      & $-$43     & \\
695  &  18 09 16.08  &  $-$23 39 59.4  & 16.40 & 0.88  & 1.91 &  0.77 &  0.25 & 12.07 &  $-$2   & ND       & $-$2    & \\
696  &  18 09 18.17  &  $-$23 34 37.7  & 16.87 & 0.95  & 2.09 &  0.81 &  0.25 & 12.38 &   ...     & ...      & $-$5:     & \\
697  &  18 09 21.94  &  $-$23 27 53.3  & ...   & ...   & ...  &  1.08 &  0.43 & 11.99 &   ...     & ...      & $-$8:     & \\
...  &  18 09 23.61  &  $-$23 27 47.9  & 14.70\tablenotemark{e} & ...   & ...  &  1.17 & 0.87  & 9.34  &   ...     & ...      & $-$168    & LkH$\alpha$ 126\\
698  &  18 09 26.04  &  $-$23 31 46.1  & 16.88 & 0.68  & 1.47 &  0.66 &   ... &  ...  &   ...     & ...      & $-$83   & \\
699  &  18 09 26.20  &  $-$23 30 09.0  & 19.95 & 1.26  & 2.90 &  1.28 &  0.75 & 12.74 &   ...     & ...      & $-$17:    & \\
700  &  18 09 26.27  &  $-$23 30 16.9  & 19.50 & 1.17  & 2.76 &  1.11 &  0.45 & 12.92 &   ...     & ...      & $-$29:    & X-ray source 17\\
701  &  18 09 26.38  &  $-$23 40 01.4  & 18.06 & 1.06  & 2.13 &  1.13 &  0.18 & 12.91 &   ...     & $-$14  & $-$12   & \\
702  &  18 09 26.41  &  $-$23 39 47.8  & 20.50 & 1.37  & 3.05 &  ...  &  ...  &  ...  &   ...     & $-$37: & ND        & \\
703  &  18 09 26.68  &  $-$23 46 01.4  & 18.27 & 1.22  & 2.67 &  0.88 &  0.34 & 12.36 &   ...     & ...      & $-$36:    & \\
704  &  18 09 27.64  &  $-$23 30 00.8  & 18.29 & 1.08  & 2.36 &  1.02 &  0.44 & 12.66 &   ...     & ...      & $-$9:     & \\
705  &  18 09 30.62  &  $-$23 36 38.5  & 16.13 & 0.85  & 1.81 &  0.92 &  0.59 & 11.41 &   ...     & $-$14  & $-$28   & \\
...  &  18 09 35.72  &  $-$23 25 19.9  & 11.81\tablenotemark{f} & ...   & ...  &  0.17 &  0.12 & 10.80 &   ...     & ...      & $-$17   & LkH$\alpha$ 127\\
706  &  18 09 42.93  &  $-$23 42 17.0  & 18.29 & 0.97  & 2.11 &  0.96 &  0.69 & 12.94 &  $-$55  & $-$67  & $-$54   & \\
707  &  18 09 43.22  &  $-$23 38 21.9  & 18.96 & 1.11  & 2.45 &  1.01 &  0.68 & 12.81 &  $-$19  & $-$52  & $-$44   & \\
708  &  18 09 44.31  &  $-$23 40 10.4  & 19.37 & 1.30  & 3.06 &  1.04 &  0.53 & 12.80 &  $-$86:   & $-$96: & $-$36   & \\
...  &  18 09 45.53  &  $-$23 38 04.1  & 17.36 & 1.13  & 2.44 &  1.05 &  0.66 & 11.41 &  $-$71  & $-$106 & $-$80   & LkH$\alpha$ 128\\
709  &  18 09 45.64  &  $-$23 41 07.1  & 18.73 & 1.14  & 2.54 &  1.12 &  0.53 & 12.73 &  $-$36:   & $-$43  & $-$86   & \\
710  &  18 09 45.98  &  $-$23 37 54.9  & 17.81 & 1.03  & 2.31 &  1.18 &  0.74 & 12.08 &  $-$21  & $-$33  &  ND       & \\
...  &  18 09 46.16  &  $-$23 38 52.4  & 16.20 & 0.85  & 1.88 &  0.88 &  0.64 & 11.29 &  $-$30  & $-$24  & $-$28   & LkH$\alpha$ 129\\
...  &  18 09 46.62  &  $-$23 38 55.1  & 16.45 & 0.83  & 2.08 &  1.00 &  0.57 & 11.09 &  $-$45  & $-$64  & $-$39   & LkH$\alpha$ 130\\ 
711  &  18 09 47.29  &  $-$23 38 41.6  & 15.87 & 0.84  & 1.84 &  0.90 &  0.53 & 11.00 &   EM      &   EM     &  EM       & \\
712  &  18 09 47.36  &  $-$23 39 16.2  & 18.86 & 1.03  & 2.32 &  1.24 &  ...  &  ...  &   EM      &   EM     & $-$30   & \\
713  &  18 09 47.75  &  $-$23 39 24.5  & 17.83 & 1.05  & 2.40 &  ...  &  ...  &  ...  &   EM      &   EM     & $-$3    & \\
714  &  18 09 47.56  &  $-$23 38 33.4  & 19.84 & 0.87  & 2.40 &  1.13 &  0.62 & 13.06 &   EM      &   EM     & $-$77:  & \\
715  &  18 09 47.84  &  $-$23 41 09.6  & 17.62 & 0.90  & 1.88 &  1.12 &  0.54 & 12.23 &   EM      & $-$33  & $-$18   & \\
716  &  18 09 48.03  &  $-$23 38 53.1  & 18.23 & 1.07  & 2.26 &  0.91 &  0.30 & 13.25 &   ND      & $-$6   & ND        & \\
717  &  18 09 48.16  &  $-$23 39 05.7  & 19.25 & 1.37  & 2.84 &  1.21 &  0.63 & 12.77 &   EM      & $-$64  & $-$119:   & \\
718  &  18 09 48.45  &  $-$23 39 12.3  & 15.09 & 0.88  & 1.92 &  1.00 &  0.55 & 10.27 &  $-$4   & $-$8   & $-$7    & \\
719  &  18 09 48.47  &  $-$23 38 18.2  & 21.02 & 1.79  & 3.66 &  0.96 &  0.48 & 13.56 &   EM      & $-$189:  & $-$113:   & \\
720  &  18 09 48.54  &  $-$23 39 14.3  & 16.57 & 0.52  & 1.89 &  ...  &  ...  & ...   &   EM      & $-$52  & $-$60:  & \\
721  &  18 09 48.50  &  $-$23 40 23.0  & 19.71 & 0.99  & 2.38 &  ...  &  0.58 & 13.48 &   EM      & $-$29: & $-$29:  & \\
722  &  18 09 49.02  &  $-$23 36 14.2  & 18.00 & 0.93  & 1.88 &  0.98 &  0.52 & 13.34 &   ND      & $-$6   &  ND       & \\
723  &  18 09 49.03  &  $-$23 38 16.4  & 15.59 & 0.90  & 1.94 &  0.82 &  0.37 & 11.02 &  $-$2   & $-$3   & $-$3    & \\
724  &  18 09 49.04  &  $-$23 38 53.4  & 18.13 & 1.10  & 2.65 &  1.08 &  0.64 & 11.70 &  $-$39: & $-$50  & $-$58   & \\
725  &  18 09 49.24  &  $-$23 41 25.2  & 19.41 & 1.30  & 2.86 &  1.10 &  0.42 & 12.67 &   EM      & $-$28: & $-$24   & \\
726  &  18 09 49.51  &  $-$23 37 07.8  & 17.50 & 1.01  & 2.21 &  0.93 &  0.35 & 12.27 &   ND      & $-$8   & $-$6    & \\
727  &  18 09 49.57  &  $-$23 38 36.6  & 20.40 & 1.40  & 2.81 &  0.98 &  0.83 & 13.49 &   EM      & $-$31: & $-$45:  & \\
728  &  18 09 49.68  &  $-$23 45 55.2  & 18.67 & 1.13  & 2.45 &  0.95 &  0.39 & 13.26 &   ...     & $-$25  &  EM       & \\
729  &  18 09 50.36  &  $-$23 38 23.0  & 19.20 & 1.29  & 2.95 &  1.00 &  0.54 & 12.80 &   EM      & $-$21: & $-$42:  & \\
730  &  18 09 50.41  &  $-$23 40 03.4  & 17.74 & 0.91  & 2.24 &  1.13 &  0.54 & 12.46 &  $-$25  & $-$28  & $-$22   & \\
731  &  18 09 50.98  &  $-$23 43 11.5  & 18.97 & 1.20  & 2.72 &  0.95 &  0.39 & 13.12 &   ...     & $-$84  & $-$94   & \\
732  &  18 09 51.10  &  $-$23 39 45.2  & 17.65 & 1.07  & 2.47 &  0.90 &  0.37 & 12.14 &  $-$14  & $-$17  & $-$23   & \\
733  &  18 09 51.32  &  $-$23 40 43.7  & 16.46 & 0.96  & 2.12 &  0.92 &  0.51 & 11.31 &  $-$10   & $-$9   & $-$7    & X-ray source 8  \\
734  &  18 09 51.82  &  $-$23 40 10.7  & 19.91 & 1.28  & 3.32 &  0.97 &  0.26 & 13.19 &   ND      & $-$49  & EM?       & \\
735  &  18 09 52.02  &  $-$23 39 05.5  & 17.00 & 0.93  & 2.05 &  1.01 &  0.38 & 11.91 &  $-$14  & $-$13  & $-$12   & \\
736  &  18 09 53.45  &  $-$23 37 33.0  & 19.40 & 1.47  & 3.21 &  1.23 &  0.90 & 11.02 &   EM      & $-$23: & EM?       & \\
737  &  18 09 54.92  &  $-$23 40 18.0  & 16.02 & 0.87  & 2.05 &  0.80 &  0.34 & 11.10 &  $-$2:    &  $-$4  & $-$4    & X-ray source 4  \\
738  &  18 09 54.99  &  $-$23 40 15.3  & 17.67 & 1.05  & 2.47 &  ...  &  ...  &  ...  &   ND      &  $-$5: & $-$8:   & \\
739  &  18 09 55.09  &  $-$23 40 26.9  & 20.15 & 1.37  & 3.31 &  0.91 &  0.54 & 12.96 &   EM      & $-$47: & $-$88:  & \\
740  &  18 09 55.09  &  $-$23 34 34.4  & 18.86 & 1.17  & 2.61 &  1.12 &  0.67 & 12.72 &   ...     & $-$14  & EM        & \\
741  &  18 09 55.71  &  $-$23 39 57.6  & 19.88 & 1.30  & 3.11 &  ...  &  0.55 & 12.96 &  EM       & $-$16: & $-$10     & \\
742  &  18 09 55.86  &  $-$23 34 35.7  & 19.97 & 1.29  & 2.79 &  ...  &  ...  &  ...  &   ...     & $-$4:  & ND        & \\
743  &  18 09 57.10  &  $-$23 31 15.7  &  ...  &  ...  & ...  &  ...  &  ...  &  ...  &   ...     &  ...     & $-$22   & \\
744  &  18 09 57.67  &  $-$23 40 01.2  & 18.51 & 1.31  & 3.05 &  1.06 &  0.55 & 11.79 &  $-$34  & $-$21  & $-$26   & X-ray source 21 \\
745  &  18 09 58.38  &  $-$23 39 15.4  & 20.06 & 1.39  & 3.28 &  0.97 &  0.48 & 13.26 &   ND    & $-$98    & $-$37   & \\
746  &  18 09 59.28  &  $-$23 37 33.8  & 21.06 & 1.54  & 3.56 &  1.66 &  1.08 & 11.36 &   ND    & $-$33:   &  EM       & \\
747  &  18 09 59.61  &  $-$23 37 29.3  & 16.95 & 1.05  & 2.36 &  1.10 &  0.71 & 11.00 &  $-$69 & $-$68   & $-$77   & \\
748  &  18 10 03.33  &  $-$23 27 47.8  &  ...  & ...   & ...  &  1.75 &  0.89 & 8.152 &   ...    &  ...      & $-$190:   & \\
749  &  18 10 02.13  &  $-$23 47 48.8  & 18.22 & 1.00  & 2.30 &  ...  &  0.80 & 10.67 &   ...    &  ...      & $-$14:    & \\
750  &  18 10 04.51  &  $-$23 39 04.5  & 18.08 & 1.04  & 2.25 &  1.10 &  0.73 & 12.11 &   ...    & $-$16   & $-$52   & \\
751  &  18 10 11.36  &  $-$23 45 06.2  & 16.26 & 0.85  & 1.90 &  0.89 &  0.56 & 11.47 &   ...    & $-$2    & ND        & \\
752  &  18 10 11.63  &  $-$23 45 37.1  & 20.31 & 1.39  & 2.90 &  1.10 &  0.58 & 13.37 &   ...    & $-$17   & EM        & \\
753  &  18 10 13.93  &  $-$23 50 19.0  & 17.26 & 0.92  & 2.26 &  0.97 &  0.47 & 11.99 &   ...    &  ...      & $-$27   & \\
754  &  18 10 15.80  &  $-$23 43 55.4  & 18.04 & 1.05  & 2.34 &  0.98 &  0.53 & 12.31 &   ...    & $-$12   & $-$10    & \\
755  &  18 10 16.00  &  $-$23 46 01.2  & 17.98 & 0.99  & 2.32 &  0.87 &  0.48 & 12.64 &   ...    & $-$34   & $-$30   & \\
756  &  18 10 22.21  &  $-$23 51 07.8  &  ...  & ...   & ...  &  1.59 &  0.70 &  9.86 &   ...    &  ...      & $-$328:   & \\
757  &  18 10 22.93  &  $-$23 51 21.3  &  ...  & ...   & ...  &   ... &  ...  &  ...  &   ...    &  ...      & $-$30   & \\
758  &  18 10 26.37  &  $-$23 50 52.0  &  ...  & ...   & ...  &  1.23 &  0.46 &  9.77 &   ...    &  ...      & $-$8    & \\	
759  &  18 10 28.33  &  $-$23 49 41.1  &  ...  & ...   & ...  &  1.10 &  1.01 & 12.95 &   ...    &  ...      & $-$30   & \\ 
760  &  18 10 30.03  &  $-$23 49 58.2  &  ...  & ...   & ...  &  1.01 &  0.76 & 11.05 &   ...    &  ...      & $-$42   & \\
761  &  18 10 36.35  &  $-$23 49 43.0  &  ...  & ...   & ...  &  1.16 &  0.61 & 12.81 &   ...    &  ...      & $-$44   & \\
762  &  18 10 43.71  &  $-$23 51 55.5  &  ...  & ...   & ...  &  0.74 &  ...  &  ...  &   ...    &  ...      & $-$146:   & \\
\enddata
\tablenotetext{a}{IH$\alpha$ number unless previously identified by Herbig (1957).}
\tablenotetext{b}{Optical photometry from the LFC, unless otherwise noted.}
\tablenotetext{c}{Near-infrared photometry from the 2MASS Point Source Catalog.}    
\tablenotetext{d}{EM: emission detected, but not measurable because of faint continuum, overlapping spectra, etc.; ND: emission not detected; ellipsis: not within survey field.}
\tablenotetext{e}{$V$-band magnitude from the literature.}
\end{deluxetable}

\begin{deluxetable}{cccccccccl}
\tabletypesize{\scriptsize}
\tablenum{4}
\tablewidth{0pt}
\tablecaption{X-ray Sources in the IC\,1274 Region}
\tablehead{
\colhead{Source}  & \colhead{$\alpha$}  & \colhead{$\delta$}  & \colhead{$V$} & \colhead{$V-R$} & \colhead{$V-I$} & \colhead{$J-H$} & \colhead{$H-K_{S}$} & \colhead{$K_{S}$} &  \colhead{Comments} \\
                &   (J2000)           &   (J2000)           &               &                 &                 &
 &                     &                   &        
}
\startdata
\phn1  &  18 09 46.74  &  $-$23 38 44.7  &  \phn8.59 & 0.18  & 0.27 &  0.17 &  0.02 &  \phn8.03 & HD 166033\tablenotemark{a} \\
\phn2  &  18 10 00.44  &  $-$23 37 25.9  & 17.97 & 0.93  & 1.91 &  0.71 &  0.17 & 13.76 &           \\
\phn3  &  18 09 51.62  &  $-$23 36 51.4  &  9.89 & 0.28  & 0.50 &  0.08 &  0.07 &  \phn9.05 & HD 314031\tablenotemark{a} \\
\phn4  &  18 08 54.92  &  $-$23 40 18.0  & 16.02 & 0.87  & 2.05 &  0.80 &  0.34 & 11.10 & IH$\alpha$-737 \\
\phn5  &  18 09 50.27  &  $-$23 32 24.6  & 21.47 & 1.86  & 3.99 &  1.30 &  0.52 & 12.62 & $\gamma$-ray source(?) \\
\phn6  &  18 09 53.89  &  $-$23 38 51.4  & 17.19 & 0.99  & 2.27 &  0.54 &  0.22 & 12.17 &            \\
\phn7  &  18 09 55.37  &  $-$23 39 54.2  & ...   &  ...  & ...  &  0.75 &  0.27 & 10.19 &            \\
\phn8  &  18 09 51.32  &  $-$23 40 43.7  & 16.46 & 0.96  & 2.12 &  0.92 &  0.51 & 11.31 & IH$\alpha$-733 \\
\phn9\tablenotemark{b}  &  18 09 49.46  &  $-$23 25 13.2  & ...   &  ...  & ...  &  1.07 &  0.89 & 10.81 & \\
10 &  18 09 59.28  &  $-$23 41 14.4  & ...   &  ...  & ...  &  1.78 &  0.92 & 13.77 & Embedded in L227\\
11 &  18 10 06.41  &  $-$23 34 56.1  & 16.11 & 0.70  & 1.45 &  0.59 &  0.13 & 12.89 &            \\
12 &  18 09 46.35  &  $-$23 40 51.5  &  \phn9.16 & 0.18  & 0.28 &  0.02 &  0.04 &  \phn8.77 & CoD $-23^{\circ}$13997\tablenotemark{a}\\
13 &  18 09 48.56  &  $-$23 39 21.1  &  \phn9.96 & 0.30  & 0.53 &  0.11 &  0.11 &  \phn8.95 & HD 314032\tablenotemark{a}  \\
14 &  18 10 00.8\phn   &  $-$23 27 56\phd\phn    & 21.64 & 1.18  & 2.67 &  ...  &  ...  &  ...  & \\
15 &  18 09 54.2\phn   &  $-$23 36 05\phd\phn    & ...   & ...   & ...  &  ...  &  ...  &  ...  & No optical or 2MASS counterpart\\
16 &  18 10 00.80  &  $-$23 38 32.4  & 17.42 & 1.20  & 2.71 &  1.06 &  0.47 & 10.84 & \\
17 &  18 09 26.27  &  $-$23 30 16.9  & 19.47 & 1.29  & 2.77 &  1.11 &  0.45 & 12.92 & IH$\alpha$-700\\
18 &  18 09 51.90  &  $-$23 38 11.6  & 16.54 & 0.94  & 2.01 &  0.87 &  0.26 & 12.01 & \\
19 &  18 09 51.46  &  $-$23 40 09.0  & 16.66 & 0.91  & 1.97 &  1.06 &  0.70 & 11.44 & \\
20\tablenotemark{b} &  18 10 52.5\phn   &  $-$23 37 13\phd\phn    & ...   & ...   & ...  &  ...  &  ...  & ...   & No 2MASS counterpart \\
21 &  18 09 57.56  &  $-$23 40 01.2  & 18.51 & 1.31  & 3.05 &  1.06 &  0.55 & 11.79 & IH$\alpha$-744\\
\enddata
\tablenotetext{a}{Optical photometry from Herbst et al.\ (1982).}
\tablenotetext{b}{Outside of LFC field of view.}
\end{deluxetable}

\begin{deluxetable}{ccccccc}
\tabletypesize{\tiny}
\tablenum{5}
\tablewidth{0pt}
\tablecaption{Derived Properties of the H$\alpha$ Emission Stars in the IC\,1274 Region}
\tablehead{
\colhead{IH$\alpha$}  & \colhead{Sp Type\tablenotemark{a}} & \colhead{$L$\tablenotemark{a}}  & \colhead{$R$\tablenotemark{a}}  & \colhead{T$_{\rm eff}$\tablenotemark{a}} & \colhead{Mass\tablenotemark{a}} & \colhead{log (Age)\tablenotemark{a}} \\
              &                   &  ($L_{\odot}$) &  ($R_{\odot}$) &         (K)         & ($M_{\odot}$)  &   (yr)        
}
\startdata
695 & K5 & 4.46 & 3.77 & 4272 & 1.02 & 5.71 \\
696 & K6 & 3.78 & 3.62 & 4096 & 0.79 & 5.67 \\
699 & K1 & 2.05 & 1.82 & 4986 & 1.55 & 6.84 \\
700 & M2 & 0.54 & 1.88 & 3484 & 0.35 & 6.11 \\
701 & M2 & 6.30 & 2.11 & 3443 & 0.34 & 6.06 \\
702 & K7 & 1.20 & 2.17 & 4056 & 0.74 & 6.16 \\
703 & M3 & 3.68 & 1.59 & 3454 & 0.34 & 6.22 \\
704 & M2 & 1.96 & 3.57 & 3569 & 0.39 & 5.00 \\ 
705 & M0 & 1.28 & 2.46 & 3797 & 0.52 & 5.90 \\
706 & K4 & 5.30 & 3.79 & 4385 & 1.24 & 5.74 \\
708 & K6 & 0.95 & 1.79 & 4077 & 0.77 & 6.29 \\
709 & M0 & 0.77 & 1.95 & 3732 & 0.48 & 6.06 \\
710 & M3 & 1.04 & 3.06 & 3310 & 0.28 & 5.03 \\
711 & K5 & 6.95 & 4.37 & 4340 & 1.22 & 5.60 \\
712 & M0 & 0.72 & 1.85 & 3826 & 0.55 & 6.16 \\
713 & M0 & 2.05 & 3.16 & 3768 & 0.50 & 5.76 \\
714 & M0 & 3.32 & 4.07 & 3739 & 0.49 & 4.85 \\
715 & M1 & 1.06 & 2.41 & 3667 & 0.44 & 5.93 \\
716 & M0 & 1.79 & 2.88 & 3833 & 0.54 & 5.81 \\
717 & K5 & 5.19 & 4.04 & 4301 & 1.10 & 5.67 \\
718 & K6 & 5.15 & 4.23 & 4106 & 0.82 & 5.58 \\
721 & M4 & 0.46 & 2.46 & 2985 & 0.19 & 5.15 \\
722 & K5 & 3.75 & 3.27 & 4291 & 1.05 & 5.79 \\
723 & M0 & 0.35 & 1.32 & 3782 & 0.52 & 6.51 \\
724 & K5 & 1.00 & 1.72 & 4301 & 1.06 & 6.53 \\
725 & K5 & 9.65 & 5.59 & 4171 & 1.01 & 5.44 \\
726 & M0 & 0.32 & 1.20 & 3768 & 0.51 & 6.55 \\
727 & M2 & 0.82 & 2.31 & 3510 & 0.36 & 5.97 \\
728 & K7 & 2.21 & 2.80 & 3950 & 0.63 & 5.82 \\
729 & M2 & 0.31 & 1.41 & 3536 & 0.37 & 6.34 \\
730 & M1 & 1.07 & 2.37 & 3699 & 0.46 & 5.94 \\
731 & M2 & 1.12 & 2.78 & 3474 & 0.35 & 5.71 \\
732 & K7 & 1.82 & 2.62 & 3924 & 0.61 & 5.84 \\
733 & M2 & 1.06 & 2.58 & 3558 & 0.38 & 5.91 \\
734 & M1 & 2.64 & 3.70 & 3718 & 0.47 & 5.13 \\
735 & K7 & 5.25 & 4.38 & 4056 & 0.77 & 5.57 \\
736 & M3 & 0.88 & 2.64 & 3381 & 0.31 & 5.64 \\
737 & K6 & 2.92 & 3.27 & 4135 & 0.82 & 5.76 \\
738 & M3 & 1.23 & 3.18 & 3410 & 0.32 & 5.07 \\
739 & K6 & 7.21 & 4.92 & 4135 & 0.91 & 5.50 \\
740 & M1 & 2.59 & 3.70 & 3718 & 0.47 & 5.18 \\
741 & M3 & 0.71 & 2.34 & 3378 & 0.31 & 5.99 \\
742 & M1 & 1.02 & 2.43 & 3623 & 0.42 & 5.94 \\
743 & M3 & 0.71 & 2.27 & 3435 & 0.33 & 6.01 \\
744 & M2 & 0.40 & 1.55 & 3573 & 0.39 & 6.24 \\
746 & M3 & 2.30 & 4.11 & 3454 & 0.34 & 4.58 \\
747 & M3 & 0.38 & 1.71 & 3391 & 0.31 & 6.19 \\
748 & M4 & 0.39 & 2.18 & 3040 & 0.20 & 5.37 \\
749 & M0 & 4.40 & 5.00 & 3797 & 0.55 & 4.73 \\
751 & M0 & 1.27 & 2.42 & 3840 & 0.55 & 5.92 \\
752 & K7 & 1.35 & 2.39 & 3898 & 0.59 & 5.93 \\
753 & K5 & 5.08 & 4.04 & 4282 & 1.07 & 5.66 \\
754 & M2 & 0.37 & 1.54 & 3503 & 0.36 & 6.24 \\
755 & K7 & 3.00 & 3.64 & 3859 & 0.57 & 5.66 \\
756 & M0 & 1.61 & 2.78 & 3797 & 0.52 & 5.83 \\
757 & K7 & 1.41 & 2.21 & 3963 & 0.65 & 6.00 \\
\enddata
\tablenotetext{a}{Predicted stellar properties from the models of Siess et al.\ (2000) for an assumed distance of 1.82 kpc and an average extinction of $A_{V}$=1.21 mag.}
\end{deluxetable}

\begin{deluxetable}{ccccccc}
\tablenum{6}
\tablewidth{0pt}
\tablecaption{Derived Properties of Optically Detected X-ray Sources in the IC\,1274 Region}
\tablehead{
\colhead{Source}  & \colhead{Sp Type\tablenotemark{a}} & \colhead{$L$\tablenotemark{a}}  & \colhead{$R$\tablenotemark{a}}  & \colhead{$T_{\rm eff}$\tablenotemark{a}} & \colhead{Mass\tablenotemark{a}} & \colhead{log (Age)\tablenotemark{a}} \\
              &                   &  ($L_{\odot}$) &  ($R_{\odot}$) &         (K)         & ($M_{\odot}$)  &   (yr)        
}
\startdata
17 & M2 &   0.68  & 2.07 & 3558 & 0.39 & 6.04 \\ 
19 & K5 &   3.68  & 3.40 & 4213 & 0.93 & 5.74 \\
18 & K6 &   4.28  & 3.60 & 4173 & 0.90 & 5.65 \\
\phn6  & M0 &   3.21  & 3.82 & 3840 & 0.56 & 5.57 \\
\phn2  & K5 &   1.05  & 1.90 & 4270 & 1.01 & 6.46 \\
14 & M1 &   0.083 & 0.70 & 3591 & 0.37 & 7.09 \\
16 & M2 &   4.24  & 4.95 & 3572 & 0.40 & 4.28 \\
11 & K1 &   4.11  & 2.55 & 5033 & 1.97 & 6.54 \\
\phn5  & M5 &   0.49  & 2.20 & 3212 & 0.25 & 5.44 \\
\enddata
\tablenotetext{a}{Predicted stellar properties from the models of Siess et al.\ (2000) for an assumed distance of 1.82 kpc and an average extinction of $A_{V}$=1.21 mag.}
\end{deluxetable}

\begin{deluxetable}{lcl}
\tablecolumns{3}
\tablenum{7}
\tablewidth{0pc}
\tablecaption{Heliocentric Radial Velocities of Early-type Stars in IC1274}
\tablehead{
\colhead{Star} & \colhead{Radial Velocity} & \colhead{UT Date} \\
    \colhead {} & \colhead{(km s$^{-1})$} & \colhead{ } 
  }
\startdata

CoD$-$23 13997 &  +12. $\pm$ 3. & 2010 Jun 4  \\
\\
  HD 166033  & \phn$-$4. $\pm$ 1. & 1997 Aug 12 \\
      &  \phn+6. $\pm$ 1. & 1998 Oct 31 \\
     &  +26. $\pm$ 1. & 2010 Jun 4 \\
 \\
  HD 314031  &  \phn+2. $\pm$ 1. & 1998 Oct 31 \\
     &  \phn+8. $\pm$ 2. & 2010 Jun 4 \\
 \\
  HD 314032  &  \phn+3. $\pm$ 2. & 2010 Jun 4 \\

\enddata
\end{deluxetable}

\begin{deluxetable}{lcccccr}
\tablecolumns{7}
\tablenum{8}
\tablewidth{0pc}
\tablecaption{ [S\ II] Lines in Cavity}
\tablehead{
\colhead{Source} & \colhead{\ $\Delta\alpha$\tablenotemark{a}}\ & \colhead{\ $\Delta\delta$\tablenotemark{a}}\ & 
\colhead{\ $\Delta$\tablenotemark{b}}\ &\colhead{} & \multicolumn{2}{r}{[S\ II]  \ \ 6717:6730  } \\ \cline{2-4} \cline{6-7}
\colhead{} &\multicolumn{3}{c}{(arcsec)} & \colhead{}  & \colhead{$v_{\sun}$\tablenotemark{c}} & \colhead{Ratio\tablenotemark{d}}\\ 
\colhead{} & \colhead{} & \colhead{} & \colhead{} & \colhead{} & \colhead{(km s$^{-1}$)} 
} 
\startdata
   T1     & 		$-$16&   		\phn+41 &  		\phn44   & &		\phn$-$5.& 		1.38  \\
   T2     & 		   +28 & 	          $-$103 & 		     107   &  &		\phn$-$5.& 		1.39 \\
   T3     & 		+96& 	\phn\phn$-$7 &  		\phn99   & &	\phn$-$3. & 1.39\\
166033 &   \phs\phn0&   \phn\phn\phs0 &   \phn\phn0  & & $-$10, $-$1  & \nodata        \\
$-$23.13997  &  \phn$-$4 &$-$126 & 126   &    &   \phn$-$5. &  1.37 \\
314032 & +27&  \phn$-$36 &  \phn45   & & \phn$-$7.  &  1.38 \\
314031 & +68&  +114 & 133   &  &  \phn$-$4.  &  1.39 \\
\enddata
\tablenotetext{a}{Rectangular coordinates of the point with respect to HD 166033 in arcsec.}
\tablenotetext{b}{Diagonal distance from  HD 166033 in arcsec.}
\tablenotetext{c}{Heliocentric velocity at the point, from the [S\ II] emission lines.}
\tablenotetext{d}{Intensity ratio of 6717/6730 at the point.}
\end{deluxetable}


\clearpage

\begin{references}
\reference{}Arias, J. I., Barba, R. H., \& Morrell, N. I. 2008, MNRAS, 374, 1253
\reference{}Basri, G. \& Marcy, G. W. 1995, AJ, 109, 762
\reference{}Bertout, C.\ 1989, \araa, 27, 351 
\reference{}Braje, T. M., Romani, R. W., Roberts, M. S. E., \& Kawai, N. 2002, ApJ, 565, L91
\reference{}Cambr{\'e}sy, L., Rho, J., Marshall, D.~J., \& Reach, W.~T.\ 2011, \aap, 527, A141 
\reference{}Carpenter, J. 2000, AJ, 120, 3139
\reference{}Crampton, D., \& Fisher, W.~A.\ 1974, Publications of the Dominion Astrophysical Observatory Victoria, 14, 283 
\reference{}Dahm, S. E. 2005, AJ, 130, 1805
\reference{}Dahm, S. E., \& Hillenbrand, L. A. 2007, AJ, 133, 2072
\reference{}Dahm, S. E., \& Simon, T. 2005, AJ, 129, 829
\reference{}Dahm, S. E., Simon, T., Proszkow, E., \& Patten, B. M.  2007, AJ, 134, 999
\reference{}Drilling, J. S., \& Landolt, A. U., in Allen's Astrophysical Quantities, ed. A. N. Cox (4th ed.; New York: Springer), 381
\reference{}Elmegreen, B. G. 2001, in ASP Conf.\ Ser.\ 243, From Darkness to Light: Origin and Evolution of Young Stellar Clusters, ed. T. Montmerle \& P. Andr\ 'e (San Francisco, CA: ASP),  255 
\reference{}Feigelson, E. D., Gaffney, J. A., III, Garmire, G., Hillenbrand, L. A., \& Townsley, L. 2003, \apj, 584, 911
\reference{}Fich, M., \& Blitz, L. 1984, 279, 125
\reference{}Georgelin, Y. M., Georgelin, Y. P., \& Roux, S. 1973, A\&A, 25, 337
\reference{}Grasdalen, G. L. 1975, PASP, 87, 831
\reference{}Haisch, K. E., Jr., Lada, E. A., \& Lada, C. J. 2001, \apjl, 553, L153 
\reference{}Hartman, R. C., Bertsch, D. L., Bloom, S. D., Chen, A. W., Deines-Jones, P., Esposito, J. A., et al.\ 1999, ApJ, 525, 191
\reference{}Hawley, S. L., Gizis, J. E., \& Reid, I. N. 1996, AJ, 112, 2799
\reference{}He, L., Whittet, D. C. B., Kilkenny, D., \& Spencer Jones, J. H. 1995, ApJS, 101, 335
\reference{}Herbig, G. H. 1957, ApJ, 125, 654
\reference{}Herbig, G. H. 1998, AJ, 497, 736
\reference{}Herbig, G. H., Andrews, S. M., \& Dahm, S. E. 2004, AJ, 128, 1233
\reference{}Herbig, G. H., \& Dahm, S. E. 2002, AJ, 123, 304
\reference{}Herbig, G. H., \& Dahm, S. E. 2006, AJ, 131, 1530
\reference{}Herbst, W., Miller, D. P., Warner, J. W., \& Herzog, A. 1982, AJ, 87, 98
\reference{}Hern{\'a}ndez, J., et al. 2007, \apj, 671, 1784
\reference{}Hillenbrand, L. A., Massey, P., Strom, S. E., \& Merrill, K. M. 1993, AJ, 106, 1906
\reference{}Hillenbrand, L. A., \& White, R. J. 2004, ApJ, 604, 741
\reference{}Hodgkin, S. T., Jameson, R. F., \& Steele, I. A. 1995, MNRAS, 274, 869
\reference{}Jaschek, C., \& Gomez, A. E., 1998, A\&A, 330, 619
\reference{}Keenan, F. P., Aller, L. H., Bell, K. L., Hyung, S., McKenna, F. C., \& Ramsbottom, C. A. 1996, MNRAS, 281, 1073
\reference{}Kenyon, S. J., \& Hartmann, L. 1995, ApJS, 101, 117
\reference{}Lada, C. J., et al.\ 2006, AJ, 131, 1574 
\reference{}Landolt, A. U. 1992, AJ, 104, 340
\reference{}Mart\'{i}n, E. L. 1998, AJ, 115, 351
\reference{}Mayne, N. J., \& Naylor, T. 2008, MNRAS, 386, 261
\reference{}Mizuno, A., Onishi, T., Yonekura, Y., Nagahama, T., Ogawa, H., \& Fukui, Y. 1995, ApJ, 445, 161
\reference{}Oka, T., Kawai, N., Naito, T., Horiuchi, T., Namiki, M., Saito, Y., Romani, R. W., \& Kifune, T. 1999, ApJ, 526, 764
\reference{}Park, B-G., Sung, H., \& Kang, Y. H. 2001, JKAS, 34, 317
\reference{}Reid, N., Hawley, S. L., \& Mateo, M. 1995, MNRAS, 272, 828
\reference{}Rho, J., Lefloch, B., Reach, W. T., \& Cernicharo, J. 2008, in ASP Monograph Publ. 5., Handbook of Star Forming Regions, Vol. 2: The Southern Sky, ed. B. Reipurth (San Francisco, CA: ASP), 509
\reference{}Salpeter, E. E. 1955, ApJ, 121, 161
\reference{}Schmidt-Kaler, T. 1982, in Landolt-Bornstein, New Series, Group 6, Vol. 2b, Stars and Star Clusters, ed. K. Schaifers \& H. H. Voight (Berlin: Springer), 10
\reference{}Scoville, N. S., Yun, M. S., Clemens, D. P., Sanders, D. B., \& Waller, W. H. 1987, ApJS, 63, 821
\reference{}Siess, L., Dufour, E., \& Forestini, M. 2000, A\&A, 358, 593
\reference{}Tokunaga, A. T. 2000, in Allen's Astrophysical Quantities, ed. A. N. Cox (4th ed.; New York: Springer), 143
\reference{}Uehara, M., et al.\ 2004, SPIE, 5492, 661
\reference{}Vogt, N., \& Moffat, A. F. J., 1975, A\&A, 45, 405
\reference{}Vogt, S. S., et al.\ 1994, SPIE, 2198, 362
\reference{}Walker, M.~F.\ 1957, \apj, 125, 636
\reference{}Walker, M.~F.\ 1959, \apj, 130, 57
\reference{}White, R. J., \& Basri, G. 2003, ApJ, 582, 1109
\reference{}Yamaguchi, N., Mizuno, N., Saito, H., Matsunaga, K., Mizuno, A., Ogawa, H., \& Fukui, Y. 1999, PASJ, 51, 791
\end{references}
\end{document}